\RequirePackage{snapshot}
\documentclass[journal]{IEEEtran}
\usepackage{amsfonts,amsmath,amstext,verbatim} 
\usepackage{amssymb}% 
\usepackage{setspace} 
\ifCLASSINFOpdf 
\usepackage[pdftex]{graphicx} 
\DeclareGraphicsExtensions{.pdf}
\usepackage[pdftex,colorlinks,%naturalnames,% 
           bookmarks=true,% 
           ]{hyperref}
\usepackage[numbers,sort&compress]{natbib} 
\usepackage{hypernat} 
\else 
\usepackage[dvips]{graphicx}
\DeclareGraphicsExtensions{.eps}
\usepackage[dvips, colorlinks,%draft, %naturalnames,% 
           bookmarks=true,%
           ]{hyperref}
 \usepackage[numbers,sort&compress]{natbib} 
\fi 

 \hypersetup{
   bookmarksnumbered,
   pdfstartview={FitH},
   citecolor={blue},
   linkcolor={red},
   urlcolor=[rgb]{0,0.55,0},%{green},
   pdfpagemode={UseOutlines},
   pdfinfo={
     Title={Random linear multihop relaying in a field of interferers
                using spatial Aloha},
     Author={B. Blaszczyszyn, P. Muhlethaler},
     Subject={},
     Keywords={}
} 
}

\newcommand{\md}{\mathrm{d}} 
\newcommand{\ir}{{\mathbb{R}}}

\newcommand{\iz}{{\mathbb{Z}}}
\newcommand{\E}{{\mathbf E}} 
\newcommand{\Pro}{{\mathbf P}} 
\newcommand{\ind}{1\hspace{-0.30em}{\mbox{I}}}

\newtheorem{Th}{Theorem}[section] 
\newtheorem{prop}[Th]{Proposition} 
\newenvironment{Prop}{\bf\begin{prop}\rm\em}{\end{prop}} % proposition 
\newtheorem{lem}[Th]{Lemma} 
\newenvironment{Lem}{\bf\begin{lem}\rm\em}{\end{lem}} % lemma 
\newtheorem{cor}[Th]{Corollary} 
\newenvironment{Cor}{\bf\begin{cor}\rm\em}{\end{cor}} % corollary 
\newtheorem{res}[Th]{Result} 
\newtheorem{fact}[Th]{Fact}
\newtheorem{Rem}[Th]{Remark} 
\newenvironment{rem}{\bf\begin{Rem}\rm}{\end{Rem}} % Numbered remark 
\newtheorem{Exe}[Th]{Example} 
\newenvironment{exe}{\bf\begin{Exe}\rm}{\end{Exe}} % Numbered remark 

\newcommand{\calC}{\,{\mathcal{C}}}  
\newcommand{\calD}{\,{\mathcal{D}}} 
\newcommand{\calG}{\,{\mathcal{G}}} 

\newcommand{\calR}{{\mathcal{R}}} 
\newcommand{\calL}{{\mathcal{L}}}

\begin{document} 
\title{Random linear multihop relaying in a general field of interferers
                using spatial Aloha}

\author{\IEEEauthorblockN{Bart{\l}omiej~B{\l}aszczyszyn\IEEEauthorrefmark{1} and 
Paul M\"uhlethaler\IEEEauthorrefmark{2}}\\
\IEEEauthorblockA{\IEEEauthorrefmark{1}Inria-ENS,
23 Avenue d'Italie, 75214 
Paris, FRANCE;
Email: Bartek.Blaszczyszyn@ens.fr\\
\IEEEauthorrefmark{2}Inria  Rocquencourt, Le Chesnay, FRANCE; 
E-mail: Paul.Muhlethaler@inria.fr}}

\maketitle 

\thispagestyle{empty} \pagestyle{empty} 
 
\begin{abstract} 
In our basic model, we study a stationary Poisson pattern of nodes on a line embedded in an independent planar Poisson field of interfering nodes.  Assuming slotted Aloha and the signal-to-interference-and-noise ratio capture condition, with the usual power-law path loss model and Rayleigh fading, we explicitly evaluate several local and end-to-end performance characteristics related to the nearest-neighbor packet relaying on this line, and study their dependence on the model parameters (the density of relaying and interfering nodes, Aloha tuning and the external noise power).  Our model can be applied in two cases: the first use is for {\em vehicular ad-hoc networks}, where vehicles are randomly located on a straight road.  The second use is to study a ``typical'' route traced in a (general) planar ad-hoc network by some routing mechanism.  The approach we have chosen allows us to quantify the non-efficiency of long-distance routing in ``pure ad-hoc'' networks and evaluate a possible remedy for it in the form of additional ``fixed'' relaying nodes, called {\em road-side units} in a vehicular network.  It also allows us to consider a more general field of interfering nodes and study the impact of the clustering of its nodes on the routing performance.  As a special case of a field with more clustering than the Poison field, we consider a {\em Poisson-line} field of interfering nodes, in which all the nodes are randomly located on random straight lines.  In this case our analysis rigorously (in the sense of Palm theory) corresponds to the typical route of this network.  The comparison to our basic model reveals a paradox: clustering of interfering nodes decreases the outage probability of a single (typical) transmission on the route, but increases the mean end-to-end delay.
\end{abstract} 
 

\section{Introduction} 
The idea of mobile ad-hoc networks (MANETs) ---
spontaneous wireless networks, operating without a backbone 
infrastructure, whose users/nodes  relay  packets for each
other  in order to enable  multihop communication ---
continues to inspire practitioners and poses interesting
theoretical questions regarding its performance capabilities. 
Vehicular ad-hoc networks (VANETs) 
may well be currently one of the most promising incarnations  of MANETs.
Promoters of VANETs believe that 
these networks will both increase safety on
the road and provide value-added services.
In order to be able to propose these 
services, numerous challenging problems must, however, be solved 
regarding efficient and robust physical layers, 
reliable and flexible medium access protocols, routing 
schemes and optimized applications. 
Using mathematical tools borrowed from stochastic geometry, 
this paper analyzes the delay 
of a packet progressing on a {\em linear route}, forwarded by randomly positioned nodes using {\em Aloha} medium access.
The packet transmissions interfere with all the nodes on the line and possibly with an {\em external} (non-participating 
in the routing) {\em field of transmitters}, both of which slow down the packet's progression.
We believe such a linear network model is adequate  
for a VANET  where vehicles are randomly located on a straight road,
and the results established in this paper shed light on VANETs' performance capabilities.

\begin{figure}[t]
\centering\includegraphics[width=1.0\linewidth,height=0.50\linewidth]{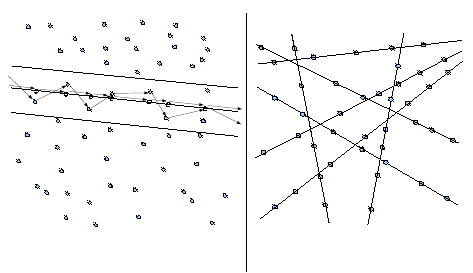}
\vspace{-3ex}
\caption[Linear relaying models]{Left : nodes relaying in given direction and in a
  given strip approximated by a line of nodes. Right: random nodes on
  random lines.}
\label{fig.lineproj}
\vspace{-4ex}
\end{figure}

We have also a second, broader, objective which is to contribute 
to the understanding  of the routing performance
in  planar (2D) MANETs.  In the context of a general 2D MANET, the  linearity (1D-pattern)
of the tagged route  may be seen as a  simplification. 
In this case, we can think of packets as being relayed in a given
direction  (e.g.  in a strip or using a geographic routing)
as depicted in the left part of
Figure~\ref{fig.lineproj}, with   the ``real''  route  approximated  by
a ``virtual'' line  joining the source and the
destination. 
%Such a situation represents an
%approximation of a geographic routing. 
We will also consider a 2D MANET with {\em all} nodes randomly  located on random  lines,
as depicted in Figure~\ref{fig.lineproj}, on the right.
Such a  2D {\em Poisson-line MANET} model (which can be formalized as a {\em doubly-stochastic
  Poisson process}, or Cox process) is an interesting alternative to traditional Poisson MANETs. 
In this case, one tagged route can be rigorously considered 
as the typical route in this Poisson-line MANET.

\subsection{Route-in-MANET model} 
More specifically, we   propose an analytically tractable multihop MANET model, by  ``decoupling'' a tagged route from the remaining 
 (external  to the route) part of the MANET. In this regard, 
we  consider two stochastically independent  point patterns to model 
respectively: some given route, on which packets of a tagged  flow are
relayed,  and  nodes which are only
sources of interference for packet transmissions on this 
route.~\footnote{In a sense, this is an extension of the well-accepted
 {\em bipolar MANET model}, in which a tagged pair 
emiter-receiver is considered in the field of interferers;
cf.~\cite[Sec.~16.2.1]{FnT2}.}
 Moreover, we assume that the tagged route is modeled by a linear
  (1D) stationary point process.
These assumptions allow us to rigorously study rigorously the progress of packets along the route.

Another advantage of the above decoupling 
is that we can go beyond Poisson assumptions regarding the distribution of nodes both on the route and 
in the interfering field.
In particular, we are able to consider  {\em Poisson nodes} on the route appended 
with a {\em periodic (lattice) structure of relaying  nodes}, which turn out to be crucial to
improve routing performance on long routes.

Regarding the external
interference, we are able to study the impact of the grouping (clustering)  of interfering nodes. In particular,
we observe and explain an interesting  paradox: more clustering in this pattern 
decreases the outage probability of a single (typical)
transmission on the route, but increases the mean end-to-end delay. 
The aforementioned 2D  Poisson-line MANET  (in which {\em all} the nodes  form independent 1D Poisson processes supported
on straight lines themselves forming a  Poisson process of lines)
is a  notable special case of a MANET which clusters nodes more than the usual Poisson MANET.
Indeed,  it has 
statistically larger empty regions 
and more numerous expected groups of points, as will be explained in this paper.

We consider slotted  Aloha MAC to be used both by the nodes
of the tagged route and of the interfering part of the MANET and 
the Signal-to-Interference-and-Noise Ratio (SINR) capture/outage
condition, with the power-law path loss model and Rayleigh fading.
We assume that all the nodes in the MANET are backlogged, i.e., they
always have packets to transmit in their buffers. Moreover,
we   are interested in the (nearest-neighbour) routing performance of a tagged
packet relayed by successive nodes on the route 
%(to their nearest neighbors)
with priority in all queues on the route.
%(cf. a discussion of these assumptions in Section~\ref{s.RW}). 
This latter assumption is particularly appropriate for safety messages, such as Decentralized Environmental Notification Messages 
(DENMs) propagated in  VANETs.

\subsection{Summary of the results} 
The main results  of this paper are the following: 
\begin{itemize} 
\item In the absence of external (to the route) noise and
  interference, we evaluate the {\em mean local delay} on the 
  Poisson route, i.e. the expected
  number of time slots required for the packet to be transmitted by
  the typical  node of the route to its nearest neighbor in  a given
  direction. The inverse of this delay is intrinsically related to the
  {\em speed of packet progression on asymptotically infinite routes}, which  can also be
  related to the {\em route transport capacity} (number of bit-meters pumped
  per unit length of the route).
\item  The mean local  delay is minimized (equivalently, the progression speed is  maximized) at a unique value of the Aloha medium access
  probability~$p$. Moreover, the routing  is unstable --- the delay is
  infinite (the speed is null) --- 
  for $p$ larger than some critical value. This observation is fully
  compliant with the phase-transition  phenomenon of the mean local delay
  discovered in the 2D Poisson MANET model in~\cite{FBBBdelay09}.  
\item We evaluate the {\em mean end-to-end delay}  of the packet
  routing between two given nodes of the route.
 % and
 % the mean packet {\em progression speed on  finite line segments of a given
 % length}. 
It  shows a cut-off phenomenon, namely that {\em
nearest-neighbor routing on distances  shorter than some critical value 
   is very inefficient in terms of exploiting
  the route transport capacity}.
\item We study the impact of noise and external interference
on our previous observations. Confirming the theoretical findings
of~\cite{BBMfppSINR} in 2D Poisson MANETs, we observe that the routing
on a Poisson route over long distances is unfeasible:
{\em the speed of packet progression on such routes is null}
due to the existence of overly large hops in random (Poisson) paths. 
Evaluating the  end-to-end delay on finite distances with noise,
we identify another,
double cut-off phenomenon: the existence of {\em critical values of
  end-to-end 
distance and noise power, beyond which the speed
of packet progression is close to~0}, again making the routing 
inefficient in terms of exploiting its  transport capacity.
\item In order to allow efficient routing over long routes
 one can complete the  Poisson route with a fixed lattice of (equidistant) 
  relaying nodes. For this model, we evaluate the mean local delay and 
  show how the route transport-capacity can  
be   maximized by an optimal choice of the inter-relay distance of the
  lattice structure.
\item We explicitly evaluate 
the   Poisson-line MANET model and compare it to the  (basic) 
Poisson-line-in-Poisson-field model.
We show that a typical transmission in the Poisson-line model 
has a smaller outage probability but the routing suffers 
larger end-to-end delays. This paradox is of a more general nature, and we  
address it in the light of the theory of clustering comparison of point processes; cf~\cite{subpoisson,dcx-clust,Soellerhaus}. 
\end{itemize}

\subsection{Related works}% and model assumptions} 
\label{s.RW} 
An almost ubiquitous stochastic assumption in the theoretical studies
of these problems in MANETs is that the
nodes of the network are distributed (at any given time) as points of a
planar (2D) Poisson point process. In conjunction with the Aloha medium access
(MAC) scheme --- simpler but less efficient than CSMA which is 
usually considered in this
context by practitioners --- 2D Poisson  MANET models allow one for quite 
explicit evaluation of several performance metrics.

Local (one-hop) characteristics in 2D Poisson MANETS,
such as SINR outage probability, the related density of packet
progress, mean local delay and many
others, have been
extensively studied in the literature; cf. e.g. among  
others~\cite{Weber-etal2005,BBM06IT,JSAC,FBBBdelay09,Haenggi2009,Haenggi2013,zzh13}.
Poisson models also allow one 
to discover some  theoretical limitations of macroscopic properties of MANETs, e.g.
the scaling  of the capacity of  dense and/or extended
networks~\cite{GuptaKumar2000,Dousse_etal2007}, or 
the speed of packet progression on long
routes~\cite{GantiHaenggi2009,BBMfppSINR}.

Obtaining quantitative, middle-scale, results (e.g. end-to-end routing delays)
is however much more difficult.   
One reason for this  is that while the source-node can be considered as a typical
node in the MANET and the powerful Palm theory of point process can be
used in the analysis of the first hop, further relay nodes on a given
path (traced by the Dijkstra algorithm or any reasonable local routing on
a 2D Poisson MANET) cannot be seen as typical nodes of a MANET.
In fact, the route followed by
a packet is a random subset of the MANET's point pattern  (depending on
the routing algorithm) and the typical point ``seen'' by the packet on a long route
is not the typical point of the whole MANET in the sense of Palm 
theory, unless a high node mobility can be assumed. This would result 
in completely independent
re-sampling of node location after each hop, which might be
reasonable in the context of delay-tolerant
networks~\cite{Jacquet-etal2010}, but which 
we do not want to assume in this paper, cf {\em routing paradox} 
in~\cite[\S~21.7.2.2]{FnT2}.

Separating (decoupling) the routing path from the remaining nodes of the network 
allows us  to consider a typical node on the route. 
More precisely, the  packets relayed through nearest-neighbor routing on a stationary, linear (1D) pattern of nodes 
``see'', at any relaying node, the typical (1D Palm) distribution of the whole route. This is a
well-known {\em point-shift invariance property of the Palm
  distribution in 1D},  which does not
have a natural extension in higher dimensions.

A very similar scenario is considered in~\cite{HaenggiSamatiou2010,net:Stamatiou11ton},
where the authors study delay and throughput in a model 
with relays which are placed equidistantly on the
source-destination line. 
Besides the fact that the topology of relaying nodes is regular (they are equidistant)
the authors of this paper assume that the pattern on interfering nodes  is  re-sampled at each slot
(in contrast to our random but static pattern).
This  major simplification allows them  
to include both the transmission delays and waiting times in the buffers on the given route (which we ignore).
 A combined TDMA/ALOHA MAC protocol with intra-route TDMA and
inter-route ALOHA is employed. 
In contrast, we use simple Aloha, ignore
queueing, and  focus on 
the performance issues caused by the {\em randomness of the topology 
of relaying nodes}. We are interested in intrinsic 
limitations of the performance 
of {\em long-distance routing} in wireless networks with an irregular
topology, which are not observable in the model with node re-sampling (cf~\cite{FBBBdelay09}).
Comparison of the fixed (correlated) and independent (re-sampled) interference
field assumption is also performed in~\cite{Crismani2013}, where  
the joint distribution of the local delays and the mean end-to-end delay on a deterministic route in Poisson fields of
interfering nodes is evaluated.

We believe that the limitations observed would
remain valid for networks employing CSMA. This is because they primarily
depend on the existence of (arbitrarily) long hops. 
CSMA, which copes better with interference, cannot improve upon 
this situation. In contrast, we show that using a regular structure of 
relay nodes superposed with irregular MANET routes leads to better 
performance. This solution is also evidently necessary for the stability of  
 queuing processes  (not covered in our  paper). 

%{\color{red} Like the present paper, \cite{zzh13} also considers local delays in 
%an Aloha-type network and characterizes thresholding effects on the mean 
%and the variance of the local delays. The difference lies in that therein the 
%hop considered has a specified length and the delays are not computed for 
%an entire route between a source and a destination. The study~\cite{zzh13} also 
%introduces MAC dynamics to reduce the effect of interference correlation which 
%is not done in the present paper.}

Regarding VANETs more specifically, \cite{bjmr12} considers the packet propagation speed
under a  protocol (not the physical, based on the SINR) 
connectivity model, assuming mobile nodes moving in opposite directions.
In this (quite different to our) scenario, 
information can either be propagated at the average speed of the vehicles or 
much faster, provided the density of vehicles moving in the 
considered propagation direction is large enough with
respect to the density of vehicles moving in the opposite direction. 
Propagation of information under protocol models is also  considered 
in~\cite{wfr04,zma10}.

{\em The remaining part of this paper} is organized as follows.
In Section~\ref{s.linearnearest}, we present our nearest neighbor routing model 
with slotted Aloha. In Section~\ref{s.deterministic} we compute routing delays 
in deterministic networks. 
In Section~\ref{s.endtoend} we study the end-to-end delays on a Poisson route
when the interference is limited to 
the interfering nodes on the Poisson route. In Section~\ref{s.noise} we  study the 
impact of external noise and interference. Section~\ref{s.conclusion} concludes 
the paper. 

\section{Linear nearest neighbor routing model in a spatial MANET with slotted Aloha} 
\label{s.linearnearest} 

\subsection{A tagged route} 
\label{ss.network} 
Let us denote by $\calR=\{X_i, i=\ldots,-1,0,1,\ldots\}$, with $X_i\in\ir^2$,
the locations of  nodes participating  in the routing of some {\em tagged flow
  of packets} from  $X_i$ to $X_{i+1}$. 
This route is assumed not to change on the time scale
considered in this paper.

The following assumptions regarding $\calR$ will be considered.

\subsubsection{Deterministic route}
Although it is not the main scenario of this paper, 
we begin by  considering a deterministic, fixed, finite pattern of nodes
 $\calR=\{X_0,\ldots,X_M\}$.

\subsubsection{Poisson-line route}
\label{ss.PL}
In this scenario we suppose that $\calR=\Phi$ forms 
a Poisson point process of intensity $\lambda$, on the line $\ir$.
In this model the notational convention is such that  $X_i< X_{i+1}$
and packets are sent by any given node $X_i$ to 
its nearest-to-the-right neighbor $X_{i+1}$. Note that the Poisson
assumption means that the 1-hop distances
$X_{i+1}-X_{i}$ are independent (across $i$) exponential random
variables with some given mean~$1/\lambda$. 
We will call this scenario {\em nearest neighbor (NN) routing on
  Poisson line}.
We will also consider  a  version of this 
model, where  the packet is transmitted to the
{\em nearest (available) receiver (NR)} to the right on $\calR$.  
This is an opportunistic routing  allowing for longer
hops, as will be explained in Section~\ref{ss.NNvsNR}.~\footnote{We implicitly 
assume that a transceiver is either in a receive (RX) or transmit (TX) 
mode, but not IDLE, nor in both receive and  transmit (RX and TX).}

\subsubsection{Poisson-line route with fixed relay nodes}
\label{sss.fixed-relays}
In this model, the tagged route consists of a superposition
$\calR=\Phi\cup\calG$, where
the Poisson route $\Phi$ is completed with 
equidistantly  located ``fixed'' relay nodes
$\calG=\{n \Delta+U_\Delta , n\in \iz\}$;  $\Delta>0$
is some fixed parameter and  $U_\Delta$ is a uniform random variable on
$[0,\Delta)$, independent of $\Phi$,  making $\calR$ stationary. 
In this model we also consider NN routing, i.e.,  packets are always 
transmitted to the  nearest neighbor to the right in $\calR$.

\subsection{A Route in a MANET}
\label{ss.RouteInManet} 
We consider  $\calR$ as some ``tagged''
 route obtained in a MANET by some routing mechanism.
A simple way of extending $\calR$ to  a 2D MANET model
consists in embedding $\calR$ in an external field of nodes $\Psi=\{Y_i\}$,
on the plane $\ir^2$. When doing so we will always consider that $\calR$ and $\Psi$ are independent.

The following assumptions regarding $\Psi$ will be considered.

\subsubsection{Fixed pattern of interferers}
One may  consider a {\em fixed, deterministic} pattern
 $\Psi=\{Y_i\}$, although, once again, it is not the main scenario
 considered in this paper.

\subsubsection{Poisson field of interferers}
\label{sss.Poisson-interf}
In this model, we assume that $\Psi$ is a Poisson point process
of some given intensity $\mu$ on $\ir^2$. The Poisson linear route
$\calR=\Phi$ embedded in a Poisson process of interferers will be our 
default {\em Poisson-line-in-Poisson field} model.

\subsubsection{Poisson-line interferers}
\label{sss.Poisson-line}
We will also consider a case where the interferers are located
on a {\em Poisson process of lines} (roads) on $\ir^2$ of rate  $\nu$
representing the total line-length  per unit of surface. 
Assuming that on each line of this process there is an independent
Poisson process 
of points of intensity $\lambda'$ nodes per unit of line length,  we obtain a 
doubly  stochastic Poisson point process on $\ir^2$ with intensity
$\mu=\lambda'\nu$ nodes per unit of surface; see~\cite{fm12}.
In particular, assuming
$\lambda=\lambda'$, one can rigorously consider the Poisson liner
route $\calR=\Phi$ embedded in such 
a Poisson line process of interferers, as the typical route of the
{\em Poisson-line MANET}, see Figure~\ref{fig.lineproj} on the right-hand 
side. 

\subsection{Aloha MAC}
We assume that all the nodes of $\calR$ and $\Psi$
try to access the channel (transmit packets) according to the Aloha 
scheme,  in  perfectly synchronized 
time slots $n=1,2,\ldots$. Each node in each time slot
independently tosses a coin with  
some bias for heads, which will be referred to as the {\em Aloha medium access probability} (Aloha MAP). Nodes whose outcome is 
heads transmit their packets, the others do not transmit. 
We denote by $p$ the MAP of nodes in $\calR$, and by $p'$ the MAP of
nodes in $\Psi$. 
The above situation will be modeled by  
marking the points $X_i\in\calR$ 
with random, Bernoulli,  medium access indicators $e_{X_i}^n$
equal to 1 for the nodes which are allowed to emit in the slot $n$ and 
0 for  the nodes which are not allowed to emit. 
We have  $\Pro(e_i^n=1)=p$ for all $i,n$.
When there is no ambiguity, we will skip the time index $n$ in the notation
and also use the notation $e_{X_i}=e_i$. Similarly, we mark the
interfering nodes by independent  Bernoulli,  medium access indicators
with parameter $p'$.
 
At each time slot $n$,
Aloha splits $\calR$ into   
two point processes  
$\calR^1=\calR^{1,n}$ of emitters (having a MAC indicator $e$ equal to 1
at time $n$) 
and (potential) receivers $\calR^0=\calR^{0,n}$. 
It is known that when $\calR=\Phi$ is a Poisson process of intensity
$\lambda$
then $\calR^1$ and $\calR^0$ are independent Poisson processes with
intensity  $\lambda p$ and
$\lambda(1-p)$, respectively.

\subsection{Signal propagation} 
\label{ss.propagation} 
Each transmitting node 
uses the same transmission power, which without loss of generality
is assumed to be equal to 1\footnote{\label{foot.power}by measuring  external  noise in ratio to the actually transmitted power}.
The signal-power path-loss  is  
modeled by the power-law function   
$l(r)=(Ar)^{\beta}$ where $A,\beta>0$ are some constants
and $r$ is the distance 
between the transmitter and the receiver.

Signal-power is also perturbed by  
random fading $F$ which is independently sampled  for each 
transmitter-receiver pair at each time slot $n$. 
Thus, the actual signal-power received at $y$ from $x$ at time $n$  
is equal to $F^n_{(x,y)}/l(|x-y|)$. 
In this paper we will restrict ourselves to  an important special case  
where $F$ is {\em exponentially 
 distributed}, which corresponds to the situation of independent 
{\em Rayleigh fading}. 
By renormalization of $A$, if required,  we can assume 
without loss of generality that $F$ has mean~1. 
 
\subsection{SINR capture} 
\label{ss.nonoutage} 
When a  node located at $x$ transmits a signal 
to a node located at $y$, 
then successful  reception depends on  
the signal-to-interference-and-noise ratio (SINR) 
\begin{equation}\label{SINR} 
\text{SINR}_{(x,y)}=\text{SINR}^n_{(x,y)}=\frac{F^n_{(x,y)}/l(|x-y|)}{W^n(y)+I_{\calR^{1,n}\setminus\{x\}}(y)}, 
\end{equation} 
where $I_{\calR^{1,n}\setminus\{x\}}$ is the shot-noise process of $\calR^{1,n}$
from which transmitter $x$ is subtracted:
$I_{\calR^{1,n}\setminus\{x\}}(y)\allowbreak=\sum_{X_i\in\calR^{1,n}\setminus\{x\}} F^n_{(y,X_i)}/l(|y-X_i|)$
representing the interference created by the nodes of route $\calR$
and $W^n(y)$ represents an external (to $\calR$) noise
(normalized by the actual transmission power, cf footnote~\ref{foot.power}). 
This external noise can be a
constant ambient noise $W^n(y)=W$ or a random field.
In particular, $W^n(y)$ may comprise the interference  
created by the external field of interferers  $\Psi$, 
 which do not belong to the
route $\calR$. In this case:
\begin{equation}\label{e.W}
W^n(y)=W+I_{\Psi^{1,n}}(y)=W+\sum_{Y_i\in\Psi^{1,n}}
F^n_{(y,Y_i)}/l(|y-Y_i|)
\, ,
\end{equation}
where  $\Psi^{1,n}$ is the subset of nodes of $\Psi$ transmitting at time~$n$.
Throughout the paper we assume that {\em  $W$, $\Psi$ and $\calR$ are mutually independent, 
with independent  MAC decisions and fading variables}.  

%In particular, in this paper 
%we will assume in this paper that the nodes of $\Psi$
%also use Aloha, with possibly different MAC $p'$.

In this paper we assume a fixed  bit-rate coding, i.e.,  
$y$ successfully receives the signal from $x$ if 
\begin{equation}\label{eq:SINR} 
\text{SINR}_{(x,y)}\ge T\,, 
\end{equation} 
where $\text{SINR}_{(x,y)}$ is given by~(\ref{SINR}) 
and $T$ is the SINR-threshold related to the bit-rate given 
some particular coding scheme.

\subsection{NN versus NR routing}
\label{ss.NNvsNR}
Recall that in our NN routing  model 
each transmitter in $\calR^1$ 
transmits to the nearest node on its right in $\calR$, 
without knowing whether it is authorized   
to transmit at this time. 
%For simplicity, we consider only slotted Aloha in this Section. 
Successful reception requires that this selected node is {\em not} 
authorized by Aloha to transmit (i.e., it is in $\calR^0$), and that the SINR 
condition~(\ref{eq:SINR}) for this transmitter-receiver pair is satisfied.  
This  corresponds to the usual separation  of the routing and MAC layers.
In contrast, NR routing consists in transmitting to the   
nearest node (in the given direction) on $\calR$ 
{\em which, at the given time slot, is not authorized by Aloha to
  transmit}. The NR model hence might be seen as 
an opportunistic  routing, which requires some interplay between MAC
and the routing layer. 
As we shall see, both routing schemes
 allow one for quite explicit performance analysis 
in our basic Poisson-line-in-Poisson-field model.

\subsection{Numerical assumptions}
\label{ss.Numass}

The default assumption in our numerical examples
throughout the paper is $\lambda=0.01$ nodes per meter in the Poisson line model, i.e. the mean 
distance between two consecutive nodes on the line is
100m. 
We will
express all (transmission) delay times in seconds (s) and 
the speed (of packet progression) in meters
{\em per slot duration}.

We will also use 
$A=1$, $\beta=4$ and $T =10$. The fading $F$ is Rayleigh, $F$ is exponentially 
distributed with mean 1, $\E(F)=1$.

\section{Preliminaries: calculus of  routing delays in deterministic networks}  
\label{s.deterministic}
Unless otherwise specified,
in this section we assume that the locations of the 
nodes in the network are known and fixed (deterministic). 
The only source of randomness
is Aloha MAC and independent Rayleigh fading between any two given nodes.
In what follows we present a simple computation which allows us to
express coverage and routing delays in such networks. Our observations
will be useful in the remaining part of this
paper,  when we will study random routes in random MANETs.

\subsubsection{Coverage probability}% and local routing delay}
\label{ss.Pi}
Let us consider a transmitter located at
$x\in \ir^2$ communicating to a receiver located at $y\in \ir^2$
in the presence of a field $\Psi=\{y_i\in\ir^2\}$ of interferers,
with all nodes subject to Aloha MAC.  For the sake of generality we
assume that $x$ and $y$ use  Aloha with  MAP~$p$, while the nodes
in $\Psi$ use a (possibly different) MAP $p'$. 
We assume that $x,y,\not\in\Psi$.
We denote by $\Pi(x,y,\Psi)$ the {\em probability of successful transmission from $x$ to $y$ in one time slot}. This probability accounts for the likelihood 
of $x$ being authorized by Aloha to transmit,  $y$ not to being
authorized to transmit and, given such circumstances, the probability
of achieving the SINR larger than $T$ at the receiver.

We define the following  functions: 
\begin{align}\label{eq.h}
h(s,r)&=1-\frac{p}{\frac{1}{T}(s/r)^{\beta}+1}\quad s,r\ge0,\\
w(s)&= \exp(-TW (As)^{\beta})\quad s\ge0.\label{eq.w}
\end{align} 
Also, we will write $h'(s,r)$ for the function given
    by~(\ref{eq.h}) with $p$ replaced by $p'$.
For reasons which will become clear in what follows,
we call $h(s,r)$ the {\em interference factor} and $w$ the {\em noise factor}.  

\begin{Lem}\label{L.Pi}
We have 
$$\Pi(x,y,\Psi) =p(1-p) w(|x-y|) \prod_{z \in \Psi} h'(|z-y|,|x-y|)\,.$$
\end{Lem}
\begin{rem}
\label{r.Pi-recur}
Let us note that $\Pi$ satisfies the following recursion
when adding an interferer $z$ to the field $\Psi$:
\begin{align}
\Pi(x,y,\emptyset) &= p(1-p)w(|x-y|)\,,\\
 \Pi(x,y,\Psi\cup\{z\})&=\Pi(x,y,\Psi)h'(|z-y|,|x-y|)\,. 
\end{align}
In other words, the external noise reduces the probability 
of successful transmission over the distance $r$
by the factor $w(r)$, which in the absence of  noise and
interference is equal
to $p(1-p)$ (because of the Aloha scheme). Moreover,  adding  an
interfering
 node $z$ to $\Psi$ causes a decrease in the successful transmission
probability  by the factor
 $h'(s,r)$, where $s$ is the distance of the new interferer to the
receiver.
%{\bf TODO successful transmission probability vs coverage probability (differ by factor $p$).}
\end{rem}
\begin{IEEEproof}[Proof of Lemma~\ref{L.Pi}]
We have
\begin{align*}
&\Pi(x,y,\{z\} )\\
&= p(1-p) \Pro\{F_{(x,y)}/l(|x-y|) \!\!\geq \!\!T(W + e_z  F_{(z,y)}/l(|y-z|)\}\\
  &= p(1-p) e^{-TWl(|x-y|)} \E\bigl[e^{-Te_z  F_{(z,y)}
    l(|x-y|)/l(|z-y|)}\bigr]
\end{align*}
and 
\begin{align}
&\E\big[e^{-Te_x  F_{(z,y)} l(|x-y|)/l(|z-y|)}\big]\\
&= (1-p')+p' \E\big[e^{-TF_{(z,y)} l(|x-y|)/l(|z-y|)}\big]\\
&=
  1-\frac{p'}{\frac1{T} \frac{|z-y|^{\beta}}{|x-y|^{\beta}}+1}\,,
\end{align}
where we use the assumption that $F$ is an exponential random variable.
The proof follows by  induction.
\end{IEEEproof}

\subsection{Local routing delay}
\label{ss.L}
$\Pi(x,y,\Psi)$ 
is the probability that node $x$
can successfully send a tagged  packet to node $y$
in a single transmission. However, a single 
transmission is not sufficient in Aloha.
Hence, after an  unsuccessful transmission, 
the transmitter will try to  retransmit the packet 
with Aloha, possibly several 
times,  until the packet's reception. 
%In the previous scenario (of $x\in \ir^2$ transmitting to 
%$y\in \ir^2$ in the presence of a field $\Psi=\{y_i\in\ir^2\}$ of
%interferers), assume that after each  unsuccessful transmission the tagged packet
%is retransmitted (from $x$ to $y$) next time when $x$ is authorized to
%transmit by Aloha. 
We denote by $L(x,y,\Psi)$ the expected number of time slots
required to successfully transmit a packet from $x$ to $y$
(considering the previous scenario of $x\in \ir^2$ transmitting to 
$y\in \ir^2$ in the presence of a field $\Psi=\{y_i\in\ir^2\}$ of
interferers).
Under our assumptions this
number is a geometric random variable of parameter  $\Pi(x,y,\Psi)$
and hence its expected value is equal to: 
$$L(x,y,\Psi):=\frac{1}{\Pi(x,y,\Psi)}\,.$$
We call $L(x,y,\Psi)$ the {\em local (routing) delay}.
By Lemma~\ref{L.Pi} we have immediately the following result.
\begin{Lem}\label{l.L}
\begin{equation}\label{e.L}
L(x,y,\Psi) =\frac1{p(1-p)} w^{-1}(|x-y|) \prod_{z \in \Psi}
h'^{-1}(|z-y|,|x-y|)\,.
\end{equation}
and an analogous recurrence to this of Remark~\ref{r.Pi-recur}
holds for $L$ with the reciprocals of the noise and interference factors. 
\end{Lem}

\subsection{Route delays}
\label{ss.Route-delays}
Let us now consider a route consisting of nodes 
$\calR=\{x_0,x_1,\dots,x_n\}$ in the field of interfering nodes  $\Psi$
(we assume that $\calR\cap\Psi=\emptyset$).  All nodes in $\calR$ use
MAP $p$ while interferers in $\Psi$ (which are considered as 
``external''to the route) use MAP $p'$.   
The average delay to send a packet from $x_0$ to $x_n$ using the route
$\calR$ is simply the sum of the local delays on successive hops 
$$L(\calR,\Psi):= \sum_{k=0}^{n-1} L(x_k,x_{k+1},\Psi \cup \calR \setminus
\{x_k,x_{k+1}\})\,.$$
Note that for the hop from $x_k$ to $x_{k+1}$ other nodes of the
route  $\calR \setminus\{x_k,x_{k+1}\}$ act as interferers.
We call $L(\calR,\Psi)$ the {\em (routing) delay} on $\calR$.%
\footnote{Observe that the routing delay  does not take into account 
packet queuing at the relay nodes.}
Using~(\ref{e.L}) we obtain
\begin{align*}
L(\calR,\Psi)=& \frac1{p(1-p)} \sum_{k=0}^{n-1}
w^{-1}(|x_k-x_{k+1}|)\\
&\times\prod_{z \in \calR \setminus \{x_k,x_{k+1}\}  }
h^{-1}(|z-x_{k+1}|,|x_k-x_{k+1}|)\\
&\times\hspace{1em} \prod_{z \in \Psi  } h'^{-1}(|z-x_{k+1}|,|x_k-x_{k+1}|)\,.
\end{align*}
We also denote  by 
$$V(\calR,\Psi):=\frac{|x_n-x_0|}{L(\calR,\Psi)}$$
the mean speed of packet progression on the route $\calR$.

\subsection{Introducing stochastic geometry}
The above expressions allow one for a relatively simple, explicit analysis 
of routing delays in fixed networks, i.e. in networks where
the locations of the nodes are fixed and known. 
In the case when such information is not
available, one adopts a stochastic-geometric approach averaging over
possible geometric scenarios regarding $\calR$ and/or $\Psi$.
For example, assume that the route $\calR$ is given and fixed,
but the number and precise locations of interferers are not given.
Assuming some statistical hypothesis regarding the
distribution of $\Psi$ (which becomes random point process)
one can calculate the average route delay 
$\E_\Psi[L(\calR,\Psi)]$ where the expectation regards the distribution of
$\Psi$. Specifically, for a given couple $(x,y)\in\ir^2$ 
we define a function
\begin{equation}
H_{x,y}(z):=\log(h(|z-y|,|x-y|)),\qquad z\in\ir^2\,
\end{equation}
and similarly $H'_{x,y}(z)$ for $h(\cdot,\cdot)$ replaced by $h'(\cdot,\cdot)$. 
We denote by $\calL_\Psi$ the Laplace transform of $\Psi$, i.e., for a
given (say non-negative) function $f(\cdot)$ defined on $\ir^2$,
$\calL_\Psi(f):=\E_\Psi\Bigl[\exp\bigl(-\sum_{Y_i\in\Psi}f(Y_i)\bigr)\Bigr]$.
We then have: 
\begin{align}\label{e.EpsiPi}
\E_\Psi[\Pi(x,y,\Psi)] &=p(1-p) w(|x-y|)\calL_\Psi(-H'_{x,y})\,,\\
\E_\Psi[L(x,y,\Psi)] &=\frac1{p(1-p)} w^{-1}(|x-y|)
\calL_\Psi(H'_{x,y})\label{e.EpsiL}
\end{align}
and 
\begin{align}\nonumber
\E_\Psi[L(\calR,\Psi)]= &\frac1{p(1-p)} \!\!\sum_{k=0}^{n-1} \!\!w^{-1}(|x_k-x_{k+1}|)
\calL_\Psi(-H'_{x_k,x_{k+1}})\\
&\times\prod_{z \in \calR \setminus \{x_k,x_{k+1}\}}\hspace{-1em}
h^{-1}(|z-x_{k+1}|,|x_k-x_{k+1}|)\,.
\label{e.EpsiR}
\end{align}

In Section~\ref{s.endtoend} we will consider a scenario when the route $\calR$
is also modeled as a point process.  Before this, let us study
the impact of the structure of the  interfering points process on the
transmission between two given points.

\subsection{More versus less clustering for the interferers}
\label{ss.Clustering}
In order to simplify the notation let us denote 
$\calL_{\Psi}^{-H'}(|x-y|):=\calL_{\Psi}(-H'_{x,y})$ and similarly 
$\calL_{\Psi}^{H'}(|x-y|):=\calL_{\Psi}(H'_{x,y})$. In what follows we
will consider two particular models for $\Psi$: Poisson point process
(P) of intensity $\mu$ (cf Section~\ref{sss.Poisson-interf}) and
Poisson-line process (PL) of line intensity $\nu$ and points-on-line
intensity $\lambda$ (cf Section~\ref{sss.Poisson-line}). For these two
processes we denote $\calL_{\Psi}^{-H'}(r)$ and $\calL_{\Psi}^{H'}(r)$
respectively by $\calL_{P}^{-H'}(r)$, $\calL_{P}^{H'}(r)$
and  $\calL_{PL}^{-H'}(r)$, $\calL_{PL}^{H'}(r)$.
\begin{Lem}\label{l.LPsiH}
For the Poisson  model of interferers $\Psi$ we have:
\begin{align}\label{e.L-HP}
\calL_{P}^{-H'}(r)&= \exp\left(-\frac{2\pi^2\mu p'
  r^2T^{2/\beta}}{\beta\sin(2\pi/\beta)}\right)\,,\\
\calL_{P}^{H'}(r)&= \exp\Bigl(\frac{2 \pi^2 \mu p'r^2 T^{2/\beta}}{ \beta (1-p')^{1-2/\beta} \sin(2 \pi/ \beta)}  \Bigr )\,.
\label{e.LHP}
\end{align}
For the  Poisson-line interferers 
we have:
\begin{align}\label{e.L-HPL}
&\calL_{PL}^{-H'}(r)= \exp\biggl( -2 \nu r T^{1/\beta} \nonumber \\
&\times \int_0^{\infty}\Bigl(1-\exp(-2\lambda'r T^{1/\beta} p'
 \int_0^{\infty} 
\frac{1}{(s^2+t^2)^{\beta/2}+1} \md t)\Bigr) \md s \biggr)\,,\\
&\calL_{PL}^{H'}(r)=\exp\biggl( -2 \nu r T^{1/\beta} \nonumber \\
&\int_0^{\infty}
\Bigl(1-\exp(2\lambda'r T^{1/\beta} p'\int_0^{\infty} 
\frac{1}{(s^2+t^2)^{\beta/2}+1-p'} \md t)\Bigr) \md s \biggr).
\label{e.LHPL}
\end{align}
\end{Lem}
\begin{IEEEproof}
All the expressions follow from the evaluation of the Laplace transforms
of the respective point processes. For a radially symmetric function
$f$ we have the following well known expression for the Laplace
transform of the Poisson process $\Psi$ of intensity $\mu$
\begin{equation}
\label{e.LT-Poisson}
\calL_{\Psi}(f)=\calL_{P}(f)=
\exp \biggl( - 2\pi \mu \int_0^{\infty}1-\exp(-f(s))
s\md s\biggr)\,.
\end{equation}
Expressions~(\ref{e.L-HP}) and~(\ref{e.LHP}) are known in the
literature, cf \cite[(16.18) and (17.34) in the context of the
  preceding equation ibid.]{FnT2}, respectively. 
The  Laplace
transform of the Poisson-line field of interferers can be found  in~\cite{fm12} 
\begin{align}
\label{e.LT-Poisson-line}
&\calL_{\Psi}(f)=\calL_{PL}(f) \nonumber\\
&=\exp \biggl( -2\nu \int_0^{\infty} 1 -e^{ -2
\lambda'\int_0^{\infty} 1-\exp(-f( \sqrt{s^2+t^2})) \md t}
\md s\!\!\biggr)\,.
\end{align}
\end{IEEEproof}

\begin{Cor}{\bf (Interferers' clustering paradox)}\label{poisson_ppl}
We have the following inequalities 
$\calL_{PL}^{-H'}(r)\ge \calL_{P}^{-H'}(r)$ and $\calL_{PL}^{H'}(r)\ge
\calL_{P}^{H'}(r)$ for all $r\ge0$, provided  $\mu=\lambda'\nu$.
In consequence, both the probability of successful transmission
$\Pi(x,y,\Psi)$ and
the local routing delay $L(x,y,\Psi)$ are larger in the model with the
Poisson-line field of
interferers than with the homogeneous Poisson field having the same 
spatial intensities of points.
\end{Cor}

The above result states that  {\em grouping interfering nodes in clusters
(on lines, in the Poisson-line model)
is beneficial for the transmission probability but harmful for the 
end-to-end delays}. This apparent paradox can be explained in the
following way. Clustered interfering nodes leaves statistically
larger ``vacant'' regions and thus increases the probability that a
given transmission from $x$ to $y$ takes place in such a
region. However, when this transmission falls within a region 
where interfering nodes have higher density,  the local delay
increases considerably and actually makes the average delay larger
even if the probability of this event (falling within a region with a high
density of interferers) is smaller. 

\begin{rem}\label{r.Laplace_ordering}
 The result of Corollary~\ref{poisson_ppl} extends to  arbitrary two point processes of
interferers $\Psi$, which are comparable in the sense of Laplace transforms 
of both positive and negative functions. More precisely,  the capture probability is expressed as the Laplace transform 
of a positive function ($-H'_{x,y}\ge0$), while the end-to-end delay requires the transform of a negative function
($H'_{x,y}\le0$). The ordering of Laplace transforms of point processes in the positive  domain is equivalent to the ordering
of void probabilities while the ordering of all moment measures implies the Laplace transform ordering in the negative
domain. It is quite natural to consider both cases as complementary ways of comparing  clustering properties of point
processes. In general, Poisson-Poisson cluster point processes (to which our Poisson-line MANET belongs)
cluster nodes more than a homogeneous Poisson process; cf~\cite{dcx-clust} and also~\cite{Soellerhaus} for more details
and numerous examples. 
\end{rem}

\begin{IEEEproof}[Proof of Corollary~\ref{poisson_ppl}]
The result follows  from the comparison of the respective Laplace
transforms of Poisson and Poisson-line processes of interferers.
This can be concluded from a more general theory of clustering comparison
of point processes, noting that our Poisson-line processes of interferers
is in fact an instance of a Poisson-Poisson cluster process, which is a
super-Poisson ($dcx$-larger than Poisson) process,
cf~\cite[Example~16]{Soellerhaus}.
The following direct comparison of the respective Laplace transforms 
confirms this general observation. Using the
inequality $e^x\ge 1+x$ for all $x$ and~(\ref{e.LT-Poisson-line}) we have:
\begin{align*}
&\calL_{PL}(f)\\
&\ge \exp \biggl( - 4 \nu \lambda' \int_0^{\infty}\int_0^\infty 1-\exp(-f( \sqrt{s^2+t^2}))
\md t \,\md s\!\!\biggr)\,\\
&\ge \exp \biggl( - 2\pi \nu \lambda' \int_0^{\infty}1-\exp(-f(s))
s\md s \!\!\biggr)\,,
\end{align*} 
with the last expression giving the Laplace transform of the
Poisson point process of intensity $\nu \lambda'$ applied to radially
symmetric function $f$. This completes the proof.
\end{IEEEproof}

\section{End-to-end delays on a Poisson route in the route-interference limited case}
\label{s.endtoend} 
In this section we assume Poisson route $\calR=\Phi$ and no external
interferers ($\Psi\equiv\emptyset$). We also assume  that 
the external noise process $W^n(y)\equiv0$ is
negligible with respect to the interference created by the nodes
participating in the routing. In this scenario we will consider NN
and NR routing schemes; cf~\ref{ss.PL}. 
We begin with a simple calculation
of the capture probability. 

\subsection{Capture probability} 
\label{ss.CpNN} 
We consider a typical node on the route $\calR$, that is of the Poisson
point process $\Phi$. By the Slivnyak theorem, it can be seen as an
``extra'' node located at the origin $X_0=0$, with the other nodes
of the route distributed according to the original stationary Poisson
process $\Phi$. All marks (Bernoulli MAC indicators, fading, etc) of
these extra nodes are independent of the marks of the points of $\Phi$.
We denote by  $\Pi_{NN}(p)$ and $\Pi_{NR}(p)$ the probability of
successful transmission of the typical node $X_0=0$ in a given
time slot to the receiver prescribed by the NN and NR routing schemes,
respectively. In the capture probability $P_{NN}(p)$ and
$P_{NR}(p)$ in the two routing schemes considered, 
we assume that the typical node is authorized to
transmit. Obviously  $\Pi_{NN}(p)=pP_{NN}(p)$ and
$\Pi_{NR}(p)=pP_{NR}(p)$.
 
 For arbitrary $a,b>0$ we denote $C(a,b)=\int_a^\infty 1/(u^b+1)\, \md u$ 
and  
\begin{equation}\label{e.C} 
C(b)=C(0,b)=\frac{\pi}{b\sin(\pi/b)}\,. 
\end{equation} 
Moreover, for given  $T$ and $\beta$ let us denote: 
\begin{align} 
\calC_1&=\calC_1(T,\beta)=T^{1/\beta}(C(T^{-1/\beta},\beta)+C(\beta))\\ 
\calC_2&=\calC_2(T,\beta)=2T^{1/\beta}C(\beta)=\frac{2T^{1/\beta}\pi}{\beta\sin(\pi/\beta)}\,. 
\end{align} 
\begin{Prop} 
\label{p.pnn} 
The probability of successful transmission by a typical node of
a Poisson route $\calR=\Phi$ authorized by Aloha to transmit to its
relay node in the NN routing model without noise 
is equal to:  
\begin{equation}\label{e.Pnnd} 
P_{NN}(p)=\frac{1-p}{1+p\calC_1 } \,
\end{equation} 
and similarly  in the NR receiver model  
\begin{equation}\label{e.Pnrd} %
P_{NR}(p)=\frac{1-p}{1+p(\calC_2-1)}\,. 
\end{equation} 
\end{Prop} 
 \begin{rem} It is easy to see that $\calC_2-1\le \calC_1$ and hence
  $P_{NR}\ge P_{NN}$, i.e., the opportunistic choice of the receiver
  pays off regarding the probability of successful transmission. 
\end{rem}
\begin{IEEEproof}[Proof of Prop.~\ref{p.pnn}] 
Note directly from the form of the SINR in~(\ref{SINR})  
that $P_{NN}(p)$ and $P_{NR}(p)$ 
do not depend on $A$. 
Hence in the remaining part of the proof we take $A=1$. 

We consider first the  NN model. 
The scenario with the typical user
$X_0=0$ located at the origin corresponds to the Palm
distribution $\Pro^0$ of the Poisson route $\calR=\Phi\cup\{0\}$.
By the known property of the Poisson point 
process, the distance $R$ from $X_0$ to its nearest neighbor to the
right, $X_1$, has an exponential distribution with 
parameter $\lambda$. Moreover given $R=r$, all  other nodes of the
Poisson route $\calR$
form a Poisson point process of intensity $\lambda$ on 
$(-\infty,0)\cup(r,\infty)$. Conditioning on $R=r$ 
we can thus use~(\ref{e.EpsiPi}) with
$\Psi=\Phi^1\cap((-\infty,0)\cup(r,\infty))=:\Phi^1_r$ 
(recall that $\Phi^1$ is a Poisson process of intensity $\lambda p$ 
of  nodes transmitting in a given time slot)
to obtain:
%the probability of the SINR condition~(\ref{eq:SINR}) is equal to  
%\begin{eqnarray*} 
%$\$P_{NN}(p)=Pro^0\{\,F \ge TI_{\Phi^1}l(r)\,|\,R=r\,\} 
%&=&\int_0^\infty e^{-s Tl(r)}\,\md \Pro(I_\Phi\le s|R=r)\\ 
%=\calL_{I_{\Phi^1}|R=r}(Tl(r))$,
%\end{eqnarray*} 
\begin{align}\nonumber
P_{NN}(p)&=\frac{\lambda}p\int_0^\infty 
e^{-\lambda r}\E^0[\Pi(0,r,\Phi^1_r)]\,\md r\\
&=\lambda (1-p)\int_0^\infty 
e^{-\lambda r}\calL_{\Phi^1_r}(H_{0,r})\,\md r\,.\label{e.PNN-cond}
\end{align}
%where  $\calL_{I_{\Phi^1}|R=r}$ denotes the Laplace transform of the   
%shot noise generated at the receiver (located at $r$) 
%by Poisson point process of intensity $p\lambda$ on 
%$(-\infty,0)\cup(r,\infty)$. 
Using the well known formula for the Laplace transform of the Poisson p.p. 
(see e.g.,~\cite[(16.4)]{FnT2}) we obtain:
%$$ 
%\calL_{I_{\Phi^1}|R=r}(\xi)= 
%\exp\Bigl\{ 
%-\lambda p\Bigl(\int_r^\infty\hspace{-1.1em}+\hspace{-0.3em}\int_0^\infty\Bigr) 
%\Bigl(1-\E[e^{-\xi F/l(s)}]\Bigr)\,\md s\Bigr\}\,. 
%$$
\begin{align*} 
\calL_{\Phi^1_r}(H_{0,r})&=\exp\Bigl\{
-\lambda p \int_{(-\infty,0)\cup(r,\infty)} 
%\int_r^\infty\hspace{-1.1em}+\hspace{-0.3em}\int_0^\infty
\Bigl(1-e^{H_{0,r}(|s|)}\Bigr)\,\md s\Bigr\}\\
&=\exp\Bigl\{-\lambda p
%\int_r^\infty\hspace{-1.1em}+\hspace{-0.3em}\int_0^\infty 
\int_{(-\infty,0)\cup(r,\infty)}
\Bigl(1-h(|s|,r)\Bigr)\,\md s\Bigr\}\\
&=\exp\Bigl\{-\lambda p
%\int_r^\infty\hspace{-1.1em}+\hspace{-0.3em}\int_0^\infty
\int_{(-\infty,0)\cup(r,\infty)}
\Bigl(\frac1{1+(|s|/r)^\beta/T}\Bigr)\,\md s\Bigr\}\\
&=\exp\Bigl\{\lambda prT^{1/\beta}(C(T^{-1/\beta}, 
\beta)+C(\beta))\Bigr\}\,.
\end{align*}
Inserting this expression in~(\ref{e.PNN-cond})
we obtain 
%\begin{eqnarray*}
%P_{NN}(p)&=&\lambda (1-p) \int_0^\infty 
%e^{-\lambda r}\exp\Bigl 
%\{-\lambda p\Bigl(\int_r^\infty\hspace{-1.1em}+
%\hspace{-0.3em}\int_0^\infty\Bigr)  \\
%&&\times \Bigl(1-\E\Bigl[e^{-l(r)TF/l(s)}\Bigr]\Bigr)\,\md s\Bigr\}\, \md r\,, 
%\end{eqnarray*}
%\end{eqnarray*} 
%where $(1-p)$ in front of the first integral is due to the fact that 
%successful reception is possible only if the nearest node is  
%not authorized to transmit at the given time slot. 
%Assuming exponential fading $F$ and the power law path-loss $l(\cdot)$ 
%we  evaluate  $1-\E[e^{-l(r)TF/l(s)}]=\frac1{1+(s/r)^\beta/T}$ and obtain  
\begin{equation}\label{e.PNN1}
P_{NN}(p)=\lambda (1-p) \int_0^\infty 
e^{-\lambda r \bigl(1+pT^{1/\beta}(C(T^{-1/\beta}, 
\beta)+C(\beta))\bigr)}\,\md r\,,
\end{equation}
which boils down to the right-hand side of~(\ref{e.Pnnd}). 

In order to obtain expression~(\ref{e.Pnrd}) for the $NR$ model, 
we follow the same process, with the following modifications. 
The distance to the nearest {\em receiver} to the right has 
the exponential distribution of parameter $\lambda(1-p)$.  
Moreover the distribution of the point process of emitters  
is Poisson with parameter $\lambda p$ and  
independent of the location of this receiver. Consequently
\begin{align}\nonumber
P_{NR}(p)&=\frac{\lambda(1-p)}p\int_0^\infty 
e^{-\lambda (1-p) r}\E^0[\Pi(0,r,\Phi^1)]\,\md r\\
&=\lambda (1-p)\int_0^\infty 
e^{-\lambda(1-p) r}\calL_{\Phi^1}(H_{0,r})\,\md r\,.\label{e.PNR}
\end{align}
Note also that by the very choice of the receiver, it is not 
authorized to emit in the given time slot, hence there is no extra $(1-p)$ factor 
in the  numerator of~(\ref{e.PNR}). We have $\calL_{\Phi^1}(H_{0,r})=\exp\{\lambda pr T^{1/\beta}2C(\beta)\}$
and plugging into~(\ref{e.PNR}) one obtains~(\ref{e.Pnrd}), which completes the proof. % 
\end{IEEEproof} 
%\begin{rem} It is easy to see that $\calC_2-1\le \calC_1$ and hence 
%  $P_{NR}\ge P_{NN}$, i.e., the opportunistic choice of the receiver 
%  pays off regarding the chance of successful transmission.  
%\end{rem} 

\subsection{Local delays}
\label{ss.delays} 

Let us denote by $L_0$ the {\em local delay} at
the tagged node in the NN routing scheme, 
i.e. the  number of consecutive time slots needed for the 
tagged node to successfully transmit a given packet to the 
receiver designated by the given routing scheme.  
Local delays will be  consider only for NN routing~\footnote{The nearest receiver is not the same across different slots. This makes the calculation of the local delay (local exit time) in NR routing
less tractable, and it is omitted in the paper due to space constraints.}.
In what follows we will give the expressions for the  
expected local delay $\E^0[L_0]$
at the typical node of the Poisson route,
which we call the
{\em average local delay}, where the  averaging 
regards  all possible Poisson  
configurations of other nodes $\Phi$.  
As we will see, this expectation is finite only if $p$ is 
sufficiently small (for any given $T$ and $\beta$).

Let us denote  
\begin{eqnarray*}
\lefteqn{\calD_1(p)=\calD_1(p;T,\beta)}\\
&=&T^{1/\beta}\Bigl( 
\int_{T^{-1/\beta}}^\infty \frac1{u^\beta+1-p}\, \md u+ 
\int_0^\infty \frac1{u^\beta+1-p}\, \md u\Bigr)\,.
\end{eqnarray*}

\begin{Prop}\label{p.emergency} 
Under the assumptions of Proposition~\ref{p.pnn}, 
the mean local delay in the NN routing model is equal to 
$$\E^0[L_0]=\frac{1}{p(1-p)(1-p\calD_1(p))}$$ 
provided  
\begin{equation}\label{delay_condition} 
p\calD_1(p)<1 
\end{equation} 
and $\overline L_{NN}=\infty$ otherwise. 
\end{Prop}

\begin{IEEEproof} 
Conditioning on the distance to the nearest neighbor as in 
the proof of Proposition~\ref{p.pnn} and using~(\ref{e.EpsiL}),
we have:
\begin{align}\nonumber
\E^0[L_0]&=\frac \lambda{p(1-p)}\int_0^\infty 
e^{-\lambda r}\calL_{\Phi^1_r}(-H_{0,r})\,\md r\,.\label{e.LNN-cond}
\end{align}
with 
\begin{align*} 
\calL_{\Phi^1_r}(-H_{0,r})&=\exp\Bigl\{
-\lambda p \int_{(-\infty,0)\cup(r,\infty)} 
%\int_r^\infty\hspace{-1.1em}+\hspace{-0.3em}\int_0^\infty
\Bigl(1-e^{-H_{0,r}(s)}\Bigr)\,\md s\Bigr\}\\
&=\exp\Bigl\{-\lambda p
%\int_r^\infty\hspace{-1.1em}+\hspace{-0.3em}\int_0^\infty 
\int_{(-\infty,0)\cup(r,\infty)}
\Bigl(1-h^{-1}(s,r)\Bigr)\,\md s\Bigr\}\\
&=\exp\{\lambda p r \calD_1(p)\}\,.
\end{align*}
Consequently, we have : 
\begin{eqnarray}
\label{e.EL}
\E^0[L_0]
%&=&\frac{\lambda}{p(1-p)}\int_0^\infty e^{-\lambda r}e^{\lambda p r 
%\calD_1(p)}\,\md r\\
&=&\frac{\lambda}{p(1-p)} 
\int_0^\infty e^{-\lambda r(1-p\calD_1(p))}\,\md r\,. 
\end{eqnarray}
It is easy to see that this integral is infinite if 
$p\calD_1(p)\ge 1$ and $1/((1-p)(1-p\calD_1(p)))$ otherwise. 
This completes the proof.
\end{IEEEproof}

\begin{rem}
Similar phase transition (existence of the critical $p$)  
for the mean local delay  was
discovered  in 2D MANETs  in~\cite{FBBBdelay09}.
Note that $\calD_1(p)\ge \calC_1 $ with the equality attained when  $p\to0$.   
Hence, for small $p$ we have  $\calD_1(p)\approx\calC_1 $, and the critical 
$p$ for which the average expected delay explodes is approximately 
$1/{\calC_1 }$.  
\end{rem}

Expression (\ref{e.EL}) can be used to tune $p$ to the
value which minimizes the mean local delay. In the next section we will see
that this value of $p$ also {\em optimizes the performance 
of NN routing on long
routes, for almost any realization of the Poisson route}.

\subsection{Long-distance speed of packet  progression }
\label{ss.Long-distance-speed}
We call long-distance speed the speed at which a packet progresses
over a long distance.
Let us denote by $D_i$, $i=1,\ldots$ the local delay of the packet 
on its hop from $X_i$ to $X_{i+1}$.
More specifically, then $\sum_{i=0}^{k-1} D_i$ is the end-to-end delay
on $k$ hops on the route that starts at $X_0=0$. Recall that 
we assume that there is no queueing delay for the packet on its path. 
We denote by $v=\lim_{k\to\infty}  |X_k|/\sum_{i=0}^k D_i$ 
the  {\em long-distance speed of the packet progression}
(expressed in the units of distance per slot duration). 
Note also that $v$ multiplied by the number of bits carried by one packet 
corresponds to the {\em route transport capacity} defined as the number of
bit-meters ``pumped'' by the network per slot.  

\begin{Prop}\label{p.ldspeed}
Under the assumptions of Proposition~\ref{p.pnn},
the mean long-distance speed of the transmission of packets in NN
routing model is equal to: 
$$v=\frac{p(1-p) (1-p\calD_1(p)) }{\lambda }$$
provided $p\calD_1(p)<1$ and 0 otherwise. 
%Similarly, in the NR model {\bf TODO}
% $$v_{NR}=...$$
\end{Prop}

\begin{IEEEproof}
 The mean empirical speed during the first $k$ hops is:
$$\frac{X_{k}}{\sum_{i=0}^{k-1} D_i}=\frac{(\sum_{i=0}^{k-1} (X_{i+1}-
  X_{i}))/k}{(\sum_{i=0}^{k-1} D_i)/k}\,   .$$
When $k$ tends 
to $\infty$ the numerator tends almost surely  to 
$1/\lambda$. In the case of Poisson (or, more generally, a renewal process) 
this follows from the classical ergodic theorem for i.i.d. random variables.
The denominator tends almost surely to $\E^0[L_0]$, which is a consequence of the ergodic theorem for marked ergodic point processes (in our case  independently marked Poisson point process); 
cf e.g.~\cite[(13.4.1 2)]{daley2}.
%{\bf TODO Assuming NR mode we have ...}
\end{IEEEproof}

%{\bf TODO: Compare numerically $v_{NN}$ and $v_{NR}$. Are they
%  similar. If so we can say that we will continue only with NN model.
\begin{exe}
\label{exe.fig1} Figure~\ref{fig.speed1} shows the mean long-distance 
speed for $\lambda=0.01$, $\beta=4$ and $T=10$. 
We observe that there is an optimal value of $p$ which maximizes 
the long-distance speed of packet progression and that this speed
drops to~0 for $p$ larger than some critical value. 
  Recall that, in terms of 
VANETs, $\lambda=0.01$ means that the averge distance between cars is
$100\text{m}$. If we assume 
that the slot duration is $1\text{ms}$, the mean long-distance 
speed is $6\text{km}/\text{s}$. An emergency packet will take $0.166\text{s}$ to
propagate over $1\text{km}$ which is approximately twice the period generally used for   
Decentralized Environmental Notification Messages (DENMs) i.e. $100\text{ms}$. 
\end{exe}

\begin{rem}\label{rem.critp}
Existence of the critical value of $p$ can be attributed to 
hops, traversed by a tagged packet on the infinite route, being
statistically too long.
 Indeed, analysing the expression in~(\ref{e.PNN1}) one sees that the ``rate'' at which
hops of length $r$ occur it is  equal to $e^{-\lambda r}$ while the
rate at which the packet passes them is $e^{-\lambda r
  pD_1(p)}$. Thus, when $pD_1(p)>1$ the packet delay becomes infinite. 
\end{rem}

\subsection{Density of progress} 
\label{ss.densityNN} 
\begin{figure}[t]
\centering\includegraphics[width=1\linewidth,height=0.5\linewidth]{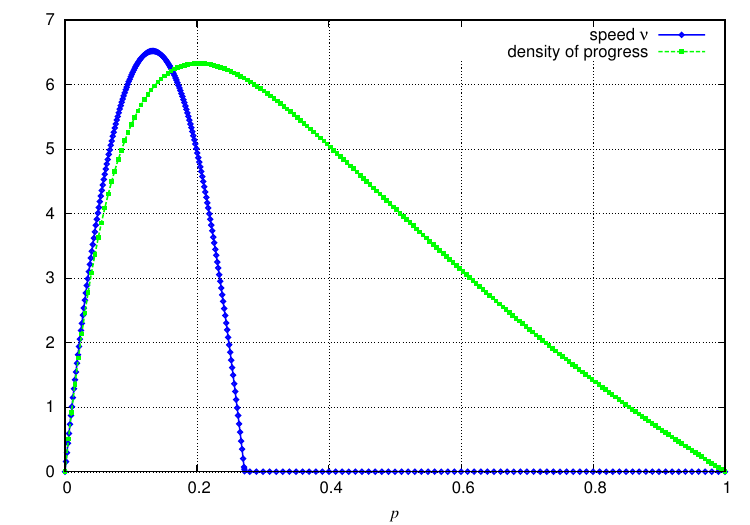}
\vspace{-3ex}
\caption[Mean speed and the  density of
  progress]{Comparison of mean speed  in the NN model and density of
  progress (both expressed in meters per slot duration); cf
  Section~\ref{ss.Numass} and Example~\ref{exe.fig1}.}
\label{fig.speed1}
\end{figure}

Some  MANET models (in particular the bipolar one)
use the  {\em mean density of progress} $d$ to optimize the
performance of the model with respect to $p$ (cf e.g.~\cite{BBM06IT}).
Recall that in a 2D MANET $d$ is defined as the  
expected total progress of all the successful transmissions  
per unit of surface. 
%In our NN model it boils down to  
%$$d=\E\Bigl[\sum_{X_i\in[0,1)} 
%|X_i-Y_{i+1}|\ind(e_{i}=1, e_{i+1}=0)\ind(\text{SINR}_{(X_i,X_{i+1})}\ge
%T)\Bigr]\,$$ 
In our 1D model, by Campbell's formula $d$ can be expressed  as: 
$$d=p(1-p)\E^0\Bigl[|X_1|\ind(\text{SINR}_{(0,X_{1})}\ge T)\Bigr]\,.$$
The density of progress, can be also seen as 
quantifying the number of bit-(or packet)-meters  
``pumped'' per unit length of a route.
\begin{Prop} 
\label{p.dnn} 
Under the assumptions of Proposition~\ref{p.pnn} 
the density of progress in 
the NN receiver model is equal to:
\begin{equation}\label{e.dnnd} 
d=d_{NN}(p)=\frac{p(1-p)}{(1+p\calC_1 )^2}\, 
\end{equation} 
and is maximized for $p$ equal to:  
$$p^*_{NN}=\frac{\calC_1 +1-\sqrt{\calC_1 ^2-1}}{2\calC_1 }\,. 
$$ 
Moreover, $0<p^*_{NN}<1$. 
%Similarly, for the NR model we have
%\begin{equation}\label{e.dnrd} 
%d_{NR}(p)=\frac{p(1-p)}{(1+p(\calC_2-1))^2}\,. 
%\end{equation} 
%and is maximized for $p$ equal to:  
%$$p^*_{NR}=\frac{\calC_2 +1-\sqrt{\calC_2 ^2-1}}{2\calC_2 }\,. 
%$$ 
%Moreover, $0<p^*_{NR}<1$. 
\end{Prop} 
\begin{IEEEproof} 
Using  Campbell's formula, the  density of progress in the NN model  
can be expressed as   
$$d_{NN}(p)=\lambda p(1-p)\E^0\Bigl[R\Pro\{\,\ind(F \ge TI_{\Phi^1}l(r)\,|\,R=r\,\}\Bigr]$$ 
with the notation as in the proof of Proposition~\ref{p.pnn}. 
Following the same arguments as in this latter  proof 
we obtain: 
$$d_{NN}(p)=\lambda^2p(1-p) \int_0^\infty 
re^{-\lambda r \bigl(1+p\calC_1 \bigr)}\,\md r\,,$$ 
which is equal to the right-hand side of~(\ref{e.dnnd}). 
Taking the derivative of the latter expression in $p$ 
we find that its sign is equal to that of 
the polynomial $P(p)=1-p(2+\calC_1 )-2p^2\calC_1 ^2+2p^3\calC_1 ^2$. 
Note that  $P(-\infty)=-\infty$, $P(0)=1$, $P(1)=-1-\calC_1 <0$ 
and $P(\infty)=\infty$.  
Hence in the 
interval $(0,1)$, $P(\cdot)$ has a unique root:  
$p^*_{NN}$ which maximizes $d_{NN}(p)$.   
The explicit expression for this root follows from  
the general formulas for the roots of cubic equations.   
%The proof for the NR model goes along the same lines.
\end{IEEEproof} 
 \begin{rem}
Note that density of progress is a ``static'' quantity, calculated
with respect to one slot. It can also be easily evaluated for the NR 
model in contrast to the mean local delay. However, it fails to discover 
the existence of the critical value of $p$ 
for the performance of the MANET revealed by the analysis of 
the local delay and packet progression speed.
 Also, one may be tempted to approximate this speed by $d/\lambda$.
Let us note, however, in Figure~\ref{fig.speed1}, 
that the optimization of this 
quantity in $p$ cannot be directly related  to the maximization of the
packet progression speed.
For this reason we will not consider the density of progress any further
in this paper.
\end{rem}

\vspace{-2ex} 

\subsection{Mean end-to-end delay
and speed of packet progression on finite segments of routes}
\label{ss.Mdelays}

\begin{figure}[h]
\centering\includegraphics[width=0.90\linewidth,height=0.13\linewidth]{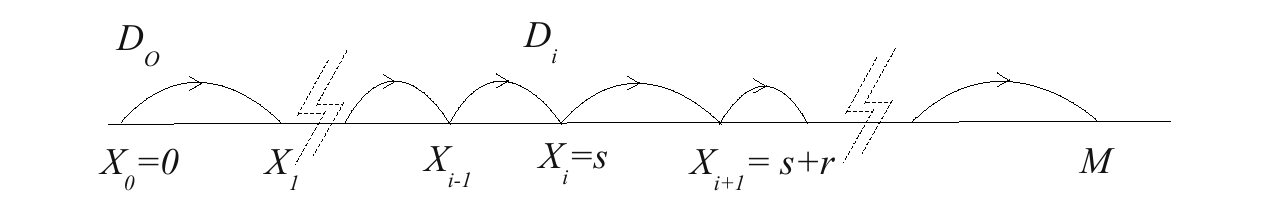}
\caption[Transmission delay on a given distance]{Transmission delay from node $0$ to node $M$.}
\label{fig.delays}
\end{figure}

In this section we will study the issues of delays and the speed of packet 
progression on finite segments of routes.
In this regard, we assume that in addition to the  node  located at the origin 
$X=0$ (also called $O$),  a second node, being the final destination
of a given packet is located at a fixed position $M>0$ on the line.
Mathematically, this corresponds to the Palm distribution $\Pro^{0M}$
of the Poisson pp, given these two fixed points. 
We want to compute 
the mean  end-to-end delay
 for a given packet to leave the node $0$ and 
reach node $M$ following NN routing.
The situation proposed is shown in Figure~\ref{fig.delays}. 
We denote this end-to-end delay by $L_{0M}=\sum_{X_i\in[0,M)}D_i$.
\vspace{-0.3 cm}

\begin{Prop}\label{p.delaysUV}
Under the assumptions of Proposition~\ref{p.pnn} 
the mean end-to-end delay in the  NN routing on the distance $M$ is
equal to
\begin{align}\label{E0M}
\lefteqn{\E^{0M}[L_{0M}]}\\
&= \frac{1}{p(1-p)} \bigg ( e^{-\lambda M} E(M)\label{a}\tag{a}\\
&\hspace{1em}+\int_0^{M} \lambda e^{-\lambda r} E(r) G_M(0,r) \md r \label{b}\tag{b}\\
&\hspace{1em} + \lambda \int_0^{M}   \int_0^{M-s} E(r) G_0(s,r)
  G_M(s,r)\lambda  e^{-\lambda r}\md r \md s \label{c}\tag{c}\\
&\hspace{1em}+ \lambda \int_0^{M} E(M-s) G_0(s,M-s) e^{-\lambda (M-s)} \md s \bigg) \label{d}\tag{d} 
\end{align}
with 
$E(r) = e^{\lambda p r \calD_1(p)}$, 
%%$$G_0(s,r)= \Big(1-\frac{p}{\frac{1}{T}  (\frac{s+r}{r})^{\beta}+1}\Big)^{-1}$$  
$G_0(s,r)= h(s+r,r)^{-1}$
and 
%%$$G_M(s,r)= \Big(1-\frac{p}{\frac{1}{T}  (\frac{M-(s+r)}{r})^{\beta}+1}\Big)^{-1}$$  
$G_M(s,r)=h(M-s-r,r)^{-1}$.
\end{Prop}
\begin{IEEEproof} 
The mean sum of the delays from the source node~0 to the destination 
node $X=M$ under $\E^{0M}$ can be expressed using Campbell's formula
 as  
$$\E^{0M}[L_{0M}]= \E^{0M}[D(0)]+ \lambda \int_{0}^M\E^{0Ms}[D(s)] \md s\,, $$
where the first term corresponds to the 
average exit time from node $0$ and $\E^{0Ms}[D(s)]$ is the average 
exit time from a ``current'' node located at $s\in(0,M)$.
These two expectations differ from $\E^{0}[L_0]$, expressed
in~(\ref{e.EL}), because there is a fixed node at $M$ which acts as an
additional interferer but which also limits the hop length. Moreover,
for the  transmission from the current node $s\in(0,M)$, 
the node at~0 also acts as an additional interferer.
We will show how the terms in~(\ref{a})--(\ref{d}) of~(\ref{E0M})
reflect these circumstances. 
Let us first remark from~(\ref{e.EL}) that 
the function $E(r)$ is the expected
exit time from a typical node, given its receiver is located at the
distance $r$ and given no additional (non-Poisson) interfering nodes.
Now, it is easy to conclude that the term~(\ref{a}) gives the expected 
end-to-end delay when it is necessary to make the 
direct hop from 0 to $M$ (when there is no Poisson relay node between them). 
The term~(\ref{b}) gives the expected exit time from 0 to its nearest
receiver when it is located at some $r\in(0,M)$:
\begin{align*}
\lefteqn{\E^{0M}[D(0);r<M]}\\
&=\frac{1}{p(1-p)}\int_0^{M} \lambda e^{-\lambda r}e^{\lambda
  p t \calD_1(p)} h(M-r,r)^{-1} \md r\,,
\end{align*}
where the factor  $h(M-r,r)^{-1}=G_M(0,r)$ (not present in~(\ref{e.EL}))
is due to the fact that $M$ acts as an additional interferer for
the transmission from 0 to $r<M$. 
Similarly, the term~(\ref{d}) corresponds to the direct hop from 
 the running node at $s$ to $M$ (when there is no Poisson relay node
 between them) and term~(\ref{c}) corresponds to the delay of going
 from $s$ to its nearest receiver $r\in(s,M)$. 
 The node at 0 interferes with both these transmissions,  which is
 reflected by the factors  $G_0(s,M-s)$ and  $G_0(s,r)$ in the 
terms~(\ref{d}) and~(\ref{c}), respectively.
The node at $M$  also interferes with the transmission from
  $s$ to  $r\in(s,M)$, whence  $G_M(s,r)$ in~(\ref{c}).
\end{IEEEproof} 
Define the  average speed of the packet progression on distance $M$
as $$v_{0M}:=M/\E^{0M}\Bigl[\sum_{X_i\in[0,M)}D_i\Bigr]=M/\E^{0M}[L_{0M}]\,.$$
Proposition~\ref{p.delaysUV} allows one to express this quantity explicitly.~\footnote{
Another  way of defining  the  average speed of the packet progression on distance $M$
consists in taking the expectation of the ratio $\E^{0M}[M/L_{0M}]$. Note (by the Jensen's inequality)
that this latter choice gives larger values of the speed, 
but both are   asymptotically equivalent in the sense that,  as $M\to\infty$,
  they approach the mean long-distance speed $v$
  considered is Proposition~\ref{p.delaysUV}. The proposed definition has an advantage of being more tractable. It is also analogous to the usual  definition of the mean user throughput describing the average speed of
data transfer in communication networks, which is 
the average number of bits sent (or
received) per data request to the average duration of the
data transfer.}
\begin{Cor}\label{p.speedUV}
The average speed of the packet progression on distance $M$
is equal to  $v_{0M}=M/\E^{0M}[L_{0M}]$ where 
$\E^{0M}[L_{0M}]$ is given by~(\ref{E0M}).
\end{Cor}
\begin{figure}[t]
\vspace{-3ex}
\centering\includegraphics[width=1\linewidth,height=0.5\linewidth]{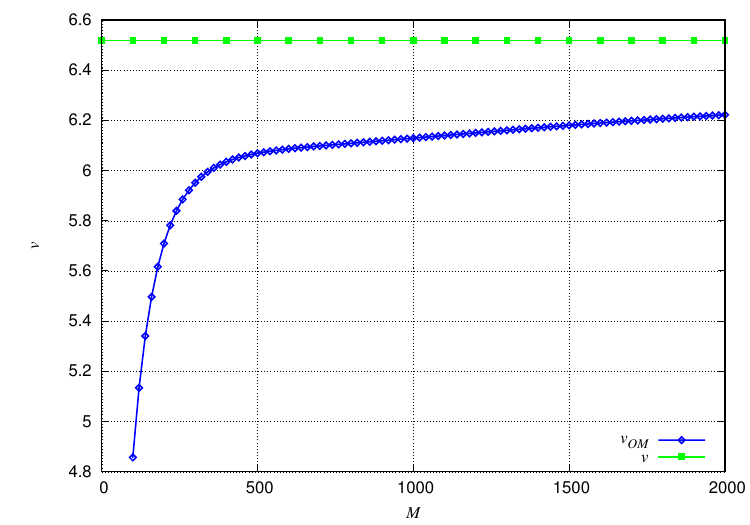}
\vspace{-3ex}
\caption[Mean short-distance speed, no noise]{Mean short-distance speed $v_{OM}$ with respect to $M$
  (expressed in meters)  in the
  NN model and long-distance speed $v$ (expressed in meters per slot duration); cf  Section~\ref{ss.Numass}.}
\label{fig.varM}
\end{figure}
%\vspace{-2ex}

\begin{exe}\label{r.flight}
In Figure~\ref{fig.varM} we present the speed of the packet progression on distance $M$
assuming $T=10$ and $\beta=4$, for $M$ varying from $100$ to
$2000$. The calculation is performed for the value of $p$ that
maximizes  the asymptotic long-distance speed calculated in
Proposition~\ref{p.ldspeed}, presented as the horizontal line 
in Figure~\ref{fig.varM}.
%\end{exe}
%
%\begin{rem}\label{r.flight}
%{\color{red} We can notice that the short-distance speed $v_{[0M)}$ 
%is always non zero whereas the long-distance speed $v$ can be zero 
%for some values of $p$, see Figure~\ref{fig.speed1}.
%}
The existence of the destination node in a finite horizon
has a double impact on the packet progression. On the one hand it ``attracts''
the packet reducing the negative impact of long hops (cf
Remark~\ref{rem.critp}).
Indeed, no hop can be longer than the direct hop to $M$ 
and thus the speed of packet progression is non-null even if $p$ is larger
than the critical value for the long-distance speed,
cf. Figure~\ref{fig.speed1}. On the other hand, it ``repels'' the packet, because it creates an additional 
interference, which is more significant when the packet is close to
the destination, cf Figure~\ref{fig.varM} with  $M$ from $0$ to $250$m  
where the speed rapidly increases with the distance $M$ to the destination. 
%for $M\lesssim 250$ and  $M\gtrsim500$ 
%the attraction (possibility of a direct
%  hop to the destination node) remains
%  strong, while the repulsion (interference created by the destination
%  node) decays fast, giving rise to a {\em fast
%    acceleration} phase. For long distances ($M\gtrsim500$), 
%the attraction is no longer  significant
%  and the repulsion decays slowly, giving rise to a {\em slow approach
%to the cruising speed}, i.e, the mean speed of packet progression on
%an infinite route. 
One can conclude that {\em routing on distances shorter than some critical value  ($M <250$m) is  inefficient in terms of exploiting of the route transport  capacity}. 
\end{exe}

\section{Impact of external noise and interference}
\label{s.noise}
In the previous section we assumed that the external noise is 
negligible ($W=0$). In this section  
we study the impact of a non-null external noise field $W^n(y)$ 
as in~(\ref{e.W}).
%More precisely, we assume that 
%Recall that we assume that the noise field  $\{W^n(y): y\in
%\ir\}$ is independent of the route process $\calR$ and that it is 
%stationary in $n$; i.e., $W^n(\cdot)$ is equal in distribution
%to $W^0(\cdot)=W(\cdot)$. We denote by $\calL_{W(r)}$ the Laplace
%transform of $W(r)$ with $r\in\ir$.
As in Section~\ref{s.deterministic}, we assume that 
the nodes of $\Psi$ use Aloha MAC with
probability $p'$. Recall also that the external noise $W$ is constant over time 
and independent both from the route $\Phi$ and the interfering nodes~$\Psi$.
 We denote 
\begin{equation}
B(r) = e^{TW(Ar)^{\beta}}\qquad
\text{and}\qquad B^{-1}(r) = e^{-TW(Ar)^{\beta}}\,. \label{e.B's}
\end{equation}

\subsection{Capture probability}
It is easy to extend the results of Propositions~\ref{p.pnn} 
and~\ref{p.dnn} to the case of an arbitrary noise field $W^n(y)$ as in~(\ref{e.W}).
\begin{Prop}\label{p.pnnW} 
The probability of successful transmission by a typical node of
Poisson route $\calR=\Phi$ authorized by Aloha to transmit to its
relay node in the NN routing model with noise and interference field $W^n(y)$ 
as in~(\ref{e.W}) 
is equal to 
\begin{equation} 
\label{e.pNN-noise}
p_{NN}=\lambda(1-p)\int_{0}^\infty 
e^{-r\lambda(1+p\calC_1)}
%\calL_{W(r)}(T(Ar)^\beta)
B^{-1}(r) \calL_{\Psi}^{-H'}(r)\,\md r\,. 
\end{equation} 
The expressions for $\calL_{\Psi}^{-H'}(r)$ in the case of
Poisson (P)  and  Poisson-line (PL) interferers are given in Lemma~\ref{l.LPsiH}.
%and the density of progress  
%\begin{equation} 
%d_{NN}=\lambda(1-p)\int_{0}^\infty 
%r e^{-r\lambda(1+p\calC_1)-TW(Ar)^\beta}\,\md r\,. 
%\end{equation} 
%Replacing $\calC_1$ by $\calC_1-1$ in the above two expressions we 
%obtain, respectively, the probability of successful transmission 
%$p_{NR}$ and the density of progress $d_{NR}$ in the NR receiver 
%model with an arbitrary noise $W$. 
\end{Prop} 
\begin{IEEEproof} 
The proof follows the same lines as in the proof of
Proposition~\ref{p.pnn},
with the averaged noise factor  $B^{-1}(r)=\calL_{W(r)}(T(Ar)^\beta)$ and the averaged 
impact of interfering field captured by $\calL_{\Psi}^{-H'}(r)$;  cf. Lemma~\ref{L.Pi}
and expression~(\ref{e.EpsiPi}).
%and Propositions~\ref{p.dnn} and~\ref{p.dnr}. 
\end{IEEEproof}

%We denote
%$C(r)=\calL_{\Psi}\Bigl(-\log h_{p'}(\cdot,r)\Bigr)$
%where  the 
%Laplace transform of the point process $\Psi$
%taken of the function $h_{p'}(\cdot,r)$ given by~(\ref{eq.h})
%with $p$  replaced by $p'$.

%\begin{Cor}\label{p.Pcexternfield}
%The  probability of successful transmission in the external field of
%interferers $\Psi$ as above
%is given by~(\ref{e.pNN-noise}) with 
%$\calL_{W(r)}(T(Ar)^\beta)$ replaced by $B^{-1}(r)\times \calL_{\Psi}^{-H'}(r)$.
%The expressions for $\calL_{\Psi}^{-H'}(r)$ in the case of
%Poisson (P)  and  Poisson-line (PL) interferers are given in Lemma~\ref{l.LPsiH}.
%\end{Cor}
%\begin{IEEEproof}
%The result follows from~(\ref{e.EpsiPi}).
%\end{IEEEproof}
%An interesting comparison of the coverage probabilities in Poisson and Poisson-line models will be
%made in Corollary~\ref{poisson_ppl}. 

\subsection{Long routes}
The following results show a very negative impact of arbitrarily
small noise on the performance of NN routing on long routes. 
\begin{Prop}\label{p.emergencyw} 
The mean local delay in the Poisson  $\calR=\Phi$ NN routing 
is infinite, $\E^0[L_0]=\infty$,
provided the noise is non-null noise,  $\Pr(W>w)\ge \epsilon$ for some
$w,\epsilon>0$, or the field of interfering nodes is super-Poisson in the sense of having  Laplace transforms of negative functions not
smaller than a Poisson point process (as e.g. Poisson-line $\Psi$)
 and applying Aloha with non-null MAP $p'>0$.
In this case the speed of packet progression on long routes is almost
surely null. 
\end{Prop} 
\begin{IEEEproof}
As in the proof of proposition~\ref{p.emergency}, cf. also Lemma~\ref{l.L}
and expression~(\ref{e.EpsiL}), we have
\begin{eqnarray*}
\E^0[L_0]\!\!\!&=&\!\!\!\frac\lambda{p(1-p)}\int_0^\infty\!\!\!\! e^{-\lambda r}e^{\lambda p r 
\calD_1(p)}B(r)\calL_\Psi^{H'}(r)\,\md r.
\end{eqnarray*} 
By our assumption on $W$, $B(r)\ge \epsilon e^{Tw(Ar)^{\beta}}$
and the integral $\int_0^\infty e^{-\lambda r(1-p\calD_1(p))+Tw(Ar)^{\beta}}\,\md r$ diverges for all $p>0$
since $\beta>2$. Regarding the impact of  $\Psi$, we consider first the Poisson case. By~(\ref{e.LHPL}),
$\calL_\Psi^{H'}(r)\sim \exp[r^2 p'/(1-p')^{1-2/\beta}\times Const]$
and consequently $\int_0^\infty e^{-\lambda r(1-p\calD_1(p))}\calL_\Psi^{H'}(r)\,\md r$ diverges for all
$p,p'>0$.  By the comparison to the Poisson case, this integral is divergent for point processes $\Psi$ which
have larger Laplace transforms of negative functions. 
\end{IEEEproof}

Recall from Remark~\ref{r.Laplace_ordering} that  larger Laplace transforms
indicate more clustering in node patterns.
\begin{rem}\label{r.Noise-vs-Inteference}
The reason why the long distance speed is $0$ lies in the fact that
the hops in a Poisson route can be arbitrarily long and that such hops
slow down (to zero) packet progression in the presence of noise.
In order to understand why even a very small noise makes  the mean local delay infinite, whereas the
in-route interference does not (for suitable small MAC $p$)
%More precisely,   the (exponential) distribution of the distance $R$
%to the nearest neighbor in 
%a Poisson pp has the tail which decreases slower than $e^{-r^\beta}$.
let us observe that this interference
(at a given receiver) varies over time because the MAC decisions and fading
variables are independent from slot to slot. Thus, even on a long hop, the packet has a chance to be transmitted when 
the in-route interference happens to be  relatively small. 
In contrast, the noise is constant (not re-sampled) from slot to slot. This makes the fundamental difference, as already observed in~\cite{FBBBdelay09}. 
Note also that external interference has the same dramatic
impact as noise, even if it also varies over time slots. 
The explanation lies in the 2D geometry
of the interfering field compared to 1D geometry of the route. More
precisely, the probability of having the nearest in-line interferer within a distance larger than $r$ is
of order $e^{-r}$, while the analogous probability regarding the nearest interferer on the plane is of
order $e^{-r^2}$ for large $r$.
 This probability is too small to let the packet traverse 
a long nearest-neighbour hop on the line, whose probability scales as  $e^{-r}$.
Let us recall from~\cite{FBBBdelay09} that the mean local delay of the transmission to the nearest neighbour in 2D
Poisson MANETs (without noise) is finite for sufficiently small $p$, since in this case both the hop-distance and the nearest
interferer distance scale as $e^{-r^2}$.
\end{rem}  

One method of coping with overly long hops consists in adding ``fixed''
points to the Poisson route $\calR$. In  fact, we will show 
in Section~\ref{ss.Short} that adding
only the fixed destination node $M$ is sufficient to make the
end-to-end delay finite in the presence of noise. 
For infinite routes one needs, however, a  ``fixed'' regular grid of relaying
nodes; cf Section~\ref{ss.lattice}.

\subsection{Short routes}
\label{ss.Short}
Our setting is now as in Section~\ref{ss.Mdelays}, i.e., 
we consider the Poisson route $\Phi$  under Palm with fixed origin and
destination nodes  $0$ and $M$, and we compute the end-to-end delay
assuming some external noise $W$ and a field of the form created by
a stationary pattern of interfering nodes $\Psi$. 

%We denote by $B(r) = e^{TW(Ar)^{\beta}}$  and
%$E'(r)=\calL_{\Psi}\Bigl(\log h_{p'}(\cdot,r)\Bigr)$ the 
%Laplace transform of the point process $\Psi$
%taken for the function $h_{p'}(\cdot,r)$ given by~(\ref{eq.h})
%with $p$  replaced by $p'$.

\begin{Prop}\label{p.delaysUV_W}
Under the assumptions of Proposition~\ref{p.pnn} and in the presence of
the external noise of the form~(\ref{e.W}), as explained above,
 the mean end-to-end delay from $0$ to $M$ in  Poisson NN routing 
 is given by~(\ref{E0M}) with 
$E(\cdot)$ (given in Proposition~\ref{p.delaysUV}) replaced by $E(\cdot)\times  B(\cdot)\times
 \calL_{\Psi}^{H'}(\cdot)$, with $B(\cdot)$ given in~(\ref{e.B's}). The expressions for $\calL_{\Psi}^{H'}(\cdot)$
 in the case of Poisson (P)  and  Poisson-line (PL) interferers are given in Lemma~\ref{l.LPsiH}.
%\begin{eqnarray*}
%\lefteqn{\E^{0M}(\sum_{X_i\in[O,M)}D_i^W)}\\
%&=& \frac{1}{p(1-p)} \bigg ( e^{-\lambda M} B(M) E(M)\\
%&&+\int_0^{M} B(r) E(r) G_M(0,r) \md r\\
%&&+ \lambda \int_0^{M}   \int_0^{M-s} B(r) E(r) G_0(s,r) G_M(s,r)\lambda  e^{-\lambda r}\md r \md s\\
%&&+ \lambda \int_0^{M} B(M-s) E(M-s) G_0(s,M-s) e^{-\lambda (M-s)} \md s \bigg)  
%\end{eqnarray*}
%with  $B(r) = e^{TW(Ar)^{\beta}}$  and $E(r),G_0(s,r),G_M(s,r)$ as defined in p%roposition~\ref{p.delaysUV}
\end{Prop}
\begin{IEEEproof}
Observe first that the conditional capture probability
given the receiver in $\Phi$ at distance $r$,
and given the interfering pattern $\Psi$, 
which we denote by $\pi(r,\Phi,\Psi )$,  
is equal to 
\begin{eqnarray*}
\lefteqn{\pi(r,\Phi, \Psi )}\\
&=& 
(1-p)\exp\Bigl\{\sum_{X_i\in\Phi}\log\Bigl(1-\frac{p}{1/T(|X_i|/r)^\beta+1}\Bigr)\Bigr\}\\
&&\times
\exp\Bigl\{\sum_{Y_i\in\Psi}\log\Bigl(1-\frac{p'}{1/T(|Y_i|/r)^\beta+1}\Bigr)\Bigr\}\\
&&\times \exp(-TW(A r)^\beta)\,;
\end{eqnarray*}
cf the proof of Proposition~\ref{p.emergency}. 
Hence, by independence, the  mean local delay, given the distance to
the receiver $R=r$, is equal to 
$$\frac{1}{p}\E^0\Bigl[\frac1{\pi(r,\Phi,\Psi)}\Big|R=r\Bigr]=
 \frac{E(r)\calL_{\Psi}^{H'}(r)B(r)}{p(1-p)}\,.$$ 
The remaining part of the proof follows the same lines as the proof of
Proposition~\ref{p.delaysUV}. 
\end{IEEEproof}

%The function $E'$ (corresponding to the Laplace transform of $\Psi$)
%can be explicitly given in the case of Poisson and Poisson-cluster  
%point processes. This  allows us to study the impact of the
%clustering of interference in a MANET. In particular, we have the
%following expressions.
%\begin{Prop}\label{p.EL-poisson-ppl}
%For Poisson field of interfering nodes of  $\Psi$ with intensity
%$\mu$ on the plane we have
% $$E'(r) =E'_{P}(r)= \exp\Bigl(\frac{2 \pi^2 \mu p' T^{2/\beta}}{ \beta (1-p')^{1-2/\beta} \sin(\frac{2 \pi}{ \beta})} r^2  \Bigr )\,.$$
%For a Poisson-line interferers (cf Section~\ref{sss.Poisson-line})
%we have 
%\begin{eqnarray*}
%E'(r)=E'_{PL}(r) &=& \exp\biggl( -2 \nu r T^{1/\beta}\int_0^{\infty}
%\Bigl(1-\exp(2\lambda'r T^{1/\beta} p'\int_0^{\infty} 
%\frac{1}{(s^2+t^2)^{\beta/2}+1-p'} \md t)\Bigr) \md s \biggr).
%\end{eqnarray*}
%\end{Prop}
%\begin{IEEEproof}
%Both expressions follow from the evaluation of the Laplace transforms
%of the respective point processes. For a Poisson process this gives 
%$$E'_{P}(r)= \exp\Bigl( 2\mu \pi r^2 T^{2/\beta}p'\int_0^{\infty}
%\frac{s}{s^{\beta}+1-p'} \md s \Bigr )$$ 
%and can be evaluated to the required expression 
%(which also appeared in~\cite[Equation
%17.43]{FnT2}. The Laplace
%transform of the Poisson-line field of interferers is given in the proof of Proposition~\ref{p.Pcexternfield} 
%from which our expression for $E'_{PL}(r)$ follows. 
%\end{IEEEproof}
%
\begin{rem}{\bf (Interferers' clustering paradox)}\label{r.poisson_ppl}
An immediate consequence of Corollary~\ref{poisson_ppl} is that 
both coverage probability $p_{NN}$ and the mean end-to-end delay
$\E^{0M}[L_{0M}]$ are larger in the model with Poisson-line field of
interferers than in the homogeneous Poisson field having the same 
spatial intensities of points $\mu=\lambda'\nu$.
Recall that this apparent paradox can be explained by noting that 
when forwarding packets in the clustering field of interferers 
one has to ``traverse'' the  regions where interfering
nodes have a higher than average density. This can considerably
slow down packet
propagation, and make it lose the advantage gained in regions with a
smaller density of interferers. The degradation increases with the
degree of clustering  (cf~\cite{dcx-clust,Soellerhaus})
 and with routing distance.
\end{rem}

\begin{figure}[t]
%\vspace{-2.8ex}
\centering\includegraphics[width=1\linewidth,height=0.45\linewidth]{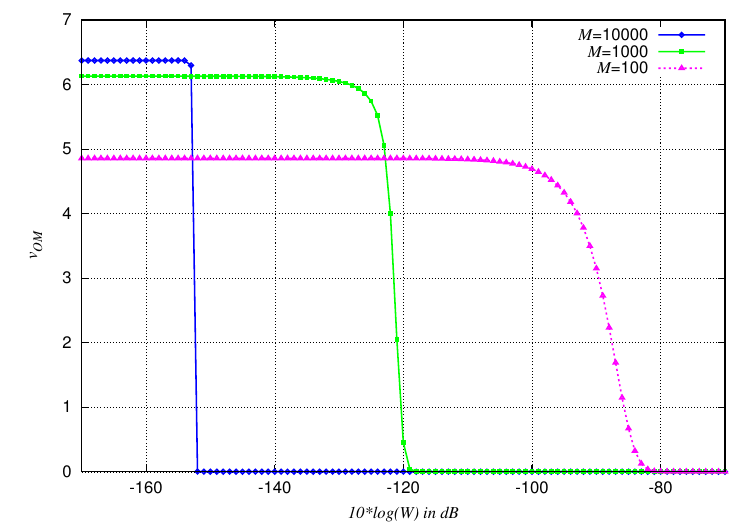}
\vspace{-4ex}
\caption[Mean short-distance speed as a function of the noise]{Mean short-distance speed (in meters per slot
  duration) with respect to noise $W$ in the NN model. 
\label{fig.varWM10000}}
\vspace{2ex}
\centering\includegraphics[width=1\linewidth,height=0.45\linewidth]{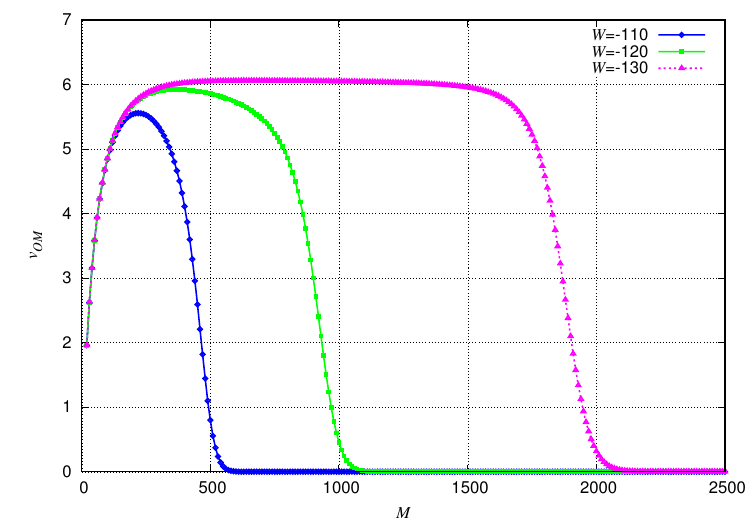}
\vspace{-4ex}
\caption[Mean short-distance speed as a function of the distance]{Mean short-distance speed  (in meters per slot
  duration) with respect to the distance $M$ (in meters) in the NN model, $W=-110,-120,-130$dB. %%$W=10^{-11},10^{-12},10^{-13}$.}
\label{fig.varMW}}
%\end{figure}
%\begin{figure}[h]
%\vspace{-1ex}
%\end{figure}
%\begin{figure}[t]
\vspace{1.5ex}
\centering\includegraphics[width=1\linewidth,height=0.45\linewidth]{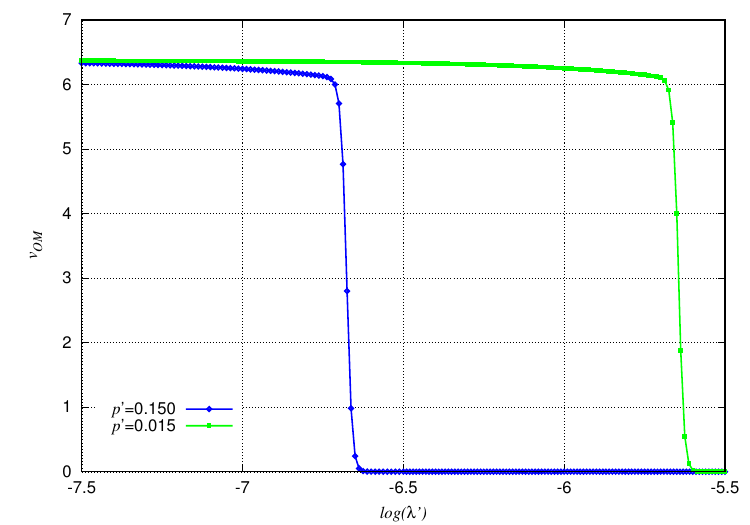}
\vspace{-4ex}
\caption[Mean short-distance speed in the Poisson field of interferers]{Mean short-distance ($M=10000$m) speed (in meters per slot
  duration) in the Poisson field of interferers with respect to it spatial intensity  $\lambda'$ (nodes per m${}^2$).
\label{fig.chlambda}}
%\vspace{-3ex}
%\end{figure}
%\begin{figure}[t]
\vspace{2ex}
\centering\includegraphics[width=1\linewidth,height=0.45\linewidth]{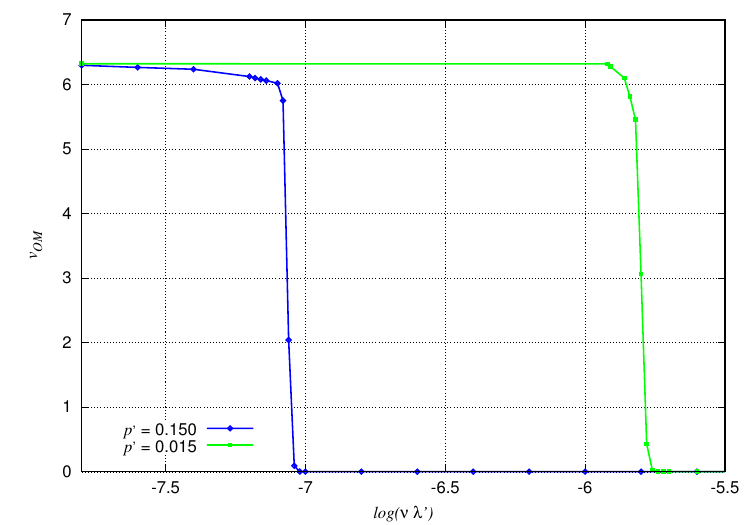}
\vspace{-4ex}
\caption[Mean short-distance speed in the  Poisson-line  field of
  interferers]{Mean short-distance ($M=10000$m) speed  (in meters per slot
  duration) in the Poisson-line of interferers, in with respect to the spatial
  intensity of nodes $\nu\lambda'$ (nodes per m${}^2$); $M=10000$m. To be compared to the Poisson case on Figure~\ref{fig.chlambda}.
\label{fig.ppl}}
%\vspace{-3ex}
\end{figure}

\begin{exe}\label{r.flightW}
The average speed of the packet progression on the distance from $O$
to $M$ can be expressed as in Corollary~\ref{p.delaysUV}.
We now present some numerical illustrations how this speed depends
on the model's parameters.
We first assume only constant noise $W$.
Figure~\ref{fig.varWM10000} shows the speed of packet progression
with respect to $W$\footnote{W is expressed in dB
according to the formula $10 \log_{10}(W/P)$ where P is the 
actually transmitted power which we suppose to be 1.}. We have assumed 
$T=10$, $\beta=4$ and $p=0.15$, the value which maximizes  
the average long-distance speed without noise.
The curves correspond to different routing distances 
 $M=100,1000,10000$. 
If we expect that an emergency packet (DENM)  be transmitted in less 
than $0.2 s$ over $1\text{km}$, we must have $W \leq -153$~dB for $M = 10000$, 
$W \leq -123$~dB for $M = 1000$. It is impossible for $M = 100$. 
Figure~\ref{fig.varMW} shows the same speed of packet progression
with respect to $M$ 
for $W=-110,-130,-150$dB.  %%$W=10^{-11},10^{-13},10^{-15}$. 
For the various values of $M$ we keep 
$p=0.15$ the value which maximizes  
the average long-distance speed without noise.
If we expect an emergency delay 
smaller than $0.2\text{s}$ over $1\text{km}$, we must have 
$M \in [120,350]$ for $W=-110$dB, 
$M \in [110,780]$ for $W=-120$dB and 
$M \in [110,1770]$ for $W=-130$dB.

%For $M=100,1000,10000$ we keep  $p=0.15$ the value which maximizes  the average long-distance speed without noise.
%% {\bf Question:  is this optimized in $p$ for each value of $M$?: Answered}
%\end{exe}
%
%\begin{rem}\label{r.flightW}
Both Figures~\ref{fig.varWM10000} and~\ref{fig.varMW}  show a sharp cut
of phenomenon: the speed of packet progression is approaching  zero 
beyond some critical noise level and/or route length.
The impact of the noise is stable below its critical value (for a given
distance).  Comparing Figure~\ref{fig.varM} to Figure~\ref{fig.varMW}
we see that after the  ``fast acceleration''
%%(cf Example~\ref{r.flightW}) 
the packet reaches 
a constant speed which can be maintained on routes shorter than some
critical length depending on the level of noise. 

Figure~\ref{fig.chlambda} shows  the 
mean end-to-end speed with respect to the spatial density 
of the field of Poisson interferers $\lambda'$
for two values of the Aloha  parameter $p'=0.15$ and $p'=0.015$
used by the  interferers. As in the case of constant noise, 
the speed rapidly falls from an almost constant value  (corresponding 
to the case without interferers) to zero. 
Figure~\ref{fig.ppl} shows the 
mean end-to-end speed with respect to the spatial density 
of the field of the Poisson-line process of rate $\nu$ and 
$\lambda'$ for $p'=0.15$ and $p'=0.015$. We observe that this 
speed is smaller than the mean speed in a field of Poisson 
interferers with a comparable spatial density 
$\nu \times \lambda'$ (cf. Figure~\ref{fig.chlambda}); this 
confirms remark~\ref{poisson_ppl}. 
For $p'=0.15$, an emergency delay over $1\text{km}$ smaller than $0.2\text{s}$  is achieved 
for $\lambda' \leq 10^{-6.7}$ in a Poisson field and for 
$\nu \lambda' \leq 10^{-7.09}$ in a Poisson process of lines. 

\end{exe}

\subsection{Equipping long Poisson routes with fixed relay stations}
\label{ss.lattice}
As previously mentioned, in order to solve the problem of long hops 
slowing down the packet progression
in the presence of external noise,
one can   add fixed relay stations, thus
allowing efficient routing on long Poisson routes. 
In this section we will study this problem. 

We assume regularly spaced ``fixed'' relay nodes
$\calG=\{n \Delta+U_\Delta , n\in \iz\}$, where $\Delta>0$ is some fixed
distance. These nodes are added to the Poisson route $\Phi$,
they use the same Aloha parameter $p$ and the  NN routing 
is considered on $\Phi\cup \calG$.

We wish to compute the mean local delay on $\Phi\cup \calG$
in the presence of noise, 
and thus characterise the  speed of packet progression on
long routes equipped with fixed relay stations.
This will allow us to optimise the performance  of the network 
in the parameter $\Delta$.  It should be noted that a small 
value of $\Delta$ makes the fixed
relay nodes too dense, creating a lot of interference and making
the routing hops very small. On the other hand, a large value of
$\Delta$ leads to long hops which slow down the packet progression.

\begin{Prop}\label{p.delaysld_W}
Assume $\calR=\Phi\cup \calG$ to be a Poisson route  equipped with fixed relay stations  with the 
inter-relay distance $\Delta$,   as described above.
The mean local delay in the NN routing on  this model in the presence
of external noise of the form~(\ref{e.W})
as in Proposition~\ref{p.delaysUV_W}
is equal to: 
\begin{eqnarray}\label{EL}
\lefteqn{\E^{0}[L]}\\
& = &\frac\lambda{p(1-p)(\epsilon+\lambda)} \bigg( \frac{1}{\Delta} \int_0^{\Delta} E(z)E'(z) e^{-\lambda z} e^{H(0,z)} \md z\nonumber\\
&&\hspace{-1em} + \frac{\lambda}{\Delta} \int_0^{\Delta} \int_0^{z}E(r)E'(z) e^{-\lambda r} e^{H(z-r,r)+\log(h(z-r,r))} \md r \md z   \bigg ) \nonumber \\
&&\hspace{-1em}+ \frac\epsilon {p(1-p)(\epsilon+\lambda)} \bigg( \frac{\lambda}{\Delta} \int_0^{\Delta} E(z)E'(z)  e^{-\lambda z} e^{H(-z,z)} \md z\nonumber\\
&&\hspace{1em} + e ^{-\lambda\Delta} B(\Delta) E(\Delta) e^{H(0,\Delta)-\log(h(\Delta,\Delta))} \bigg )\nonumber
\end{eqnarray}
with $\epsilon=\frac1{\Delta}$ and 
$H(z,r) = \sum_{n \in \iz, n \neq 0}\log(h(n\Delta+z,r))$.
\end{Prop}
\begin{IEEEproof}
%{\bf To be revised}
Let us suppose that a typical node 0 of $\Psi \cup \calG$ is at $X=0$. We know that this 
typical node belongs to $\Psi$ with probability $\frac\lambda{\epsilon+\lambda}$ and 
to $\calG$ with probability $\frac\epsilon{\epsilon+\lambda}$. Thus we have 
$$\E^{0}[L]=\frac\lambda{\epsilon+\lambda} \E^{0}[L|0\in \Psi]+ \frac\epsilon{\epsilon+\lambda} \E^{0}[L|0\in \calG].$$ 
Let us now assume that the typical node $0$ is in $\Psi$ and we use the fact that the 
nodes in $\Psi$ are independent of the nodes in $\calG$. If its closest node (towards the right) is in $\calG$, the average delay to leave node $0$ is : 
$ B(z) E(z) e^{H(0,z)}$. Conditioning by the fact there is no node of 
$\Psi$ in $[0,z]$
(that occurs with probability $e^{-\lambda z}$), we find the first integral in 
the first line of~(\ref{EL}). We 
still assume that 
the typical node is in $\Psi$ but now we also assume that the closest 
node (towards the right) is in $\Psi$. Computing the average delay in 
this case, we find the second double integral of the first line of~(\ref{EL}); 
$\lambda e^{-\lambda r}$ is actually the density of the right-hand neighbor's position. 
The contribution $e^{\log(h(z-r,r))}$ corresponds to the fact that the node in $\calG $ located 
at $z$ is not taken into account by $e^{H(z-r,r)}$  since the summation in $H$ is  
for $n \neq  0$. 

The second part of the contribution for $\E^{0}[L]$ corresponds to the case where node $0$ is in $\calG$. If we assume that the right-hand neighbor of $0$ is in $\Psi$, the 
contribution of the delay corresponds to the integral of the proposition in 
the second line of $\E^{0}[L]$. There is no correction needed for $H(-z,z)$ since the 
typical node $0$ is in $\calG$ and does not contribute to the delay as it is in $H(-z,z)$. 
The last term of the second line in~(\ref{EL}) 
corresponds to 
the case where node $0$ and its right-hand neighbor are in $\calG$. This occurs with 
probability $e^{-\lambda \Delta}$. The node in $\calG$ located at $\Delta$ does 
not contribute to the delay since we analyze the delay from $0$ and its right-hand 
neighbor in $\calG$ 
(thus located at $\Delta$).
Thus $H(0,\Delta)$ must be corrected by $e^{-\log(h(\Delta,\Delta))}$.
\end{IEEEproof}

\begin{figure}[t]
\vspace{-2ex}
\centering\includegraphics[width=1\linewidth,height=0.5\linewidth]{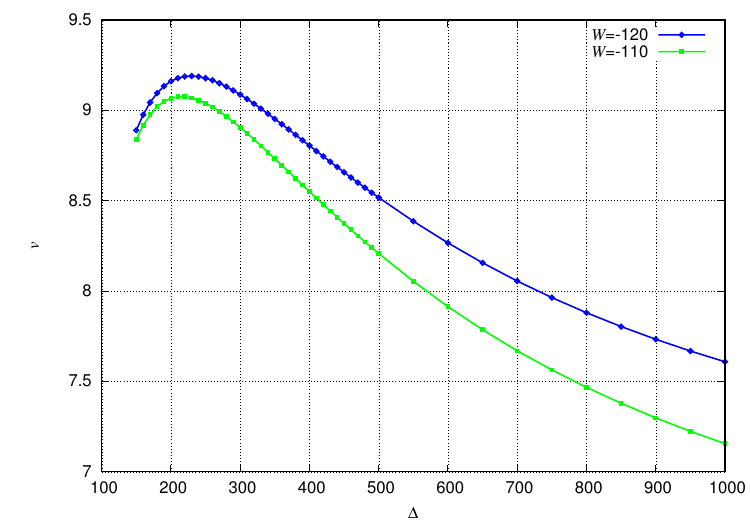}
\vspace{-4ex}
\caption[Mean long-distance speed with lattice relays]{Mean long-distance speed (in meters per slot
  duration) with respect to $\Delta$ (in meters) for  $W=-110$dB and  $W=-120$dB.} %%$W=10^{-11}$ and $W=10^{-12}$. }
\label{fig.grw}
\vspace{-1ex}
\end{figure}

\begin{exe}
In Figure~\ref{fig.grw}, we plot the mean long-distance speed 
$v=1/(\lambda \E^{0}[L])$ with respect to $\Delta$ 
with fixed noise of levels $W=-110$dB and  $W=-120$dB.  %$W=10^{-11}$ and $W=10^{-12}$. 
%\end{exe}
%
%\begin{rem}
Comparing Figure~\ref{fig.varMW}  to Figure~\ref{fig.grw} one can   
observe that the optimal inter-relay distance $\Delta$ of the fixed 
structure corresponds to the smallest length  $M$ of the route on which 
the mean speed of packet progression attains its maximum
 without the fixed infrastructure. In other words, the relay stations
 should ``break down'' long routes into segments of the length 
optimal for the end-to-end routing given the noise level. 
The emergency delay for $1 \text{km}$ is $0.103$s for
  $W=-110$dB and  $0.108$  for $W=-120$dB. 
\end{exe}

%\subsection{General summary of the results}
%\label{ss.Summary}
%In table~\ref{table.ln}, we summarize the results of a message 
%propagation in a linear network. Our model corresponds to the case 
%of a road with no other nodes than the vehicles which are on it. Without noise,
%the long-distance speed is constant. If we measure the 
%propagation speed between two fixed vehicles, the short-distance propagation speed
%approaches the long-distance speed when the fixed vehicles are 
%far apart; this speed quickly decreases for small values of the distance. 
%With noise the long-distance speed is zero, in other words a packet is 
%not propagated over very long distances. Short-distance propagation speed is 
%nearly constant when the distance between the two fixed vehicle within
%a certain 
%interval; outside this interval the short-distance propagation speed 
%is nearly zero. 

%In table~\ref{table.lni}, we summarize the long- and short-distance propagation speed for 
%packet propagating on a linear network immersed in a field of interference. We consider a 
%2D Poisson Point Process or a Poisson Process of lines. In both situations the long-distance speed 
%is zero, thus a packet is not propagated over very long paths. For
%short distances, the speed 
%is nearly constant up to a given density of interferers and rapidly approaches zero 
%for greater densities of interferers. The cut-off value of the density is larger
%for the Poisson field of interferers than for the Poisson process of lines. 
%
%
%Bla bla bla
%
%Relay station...

\section{Concluding Remarks}
\label{s.conclusion}

\begin{table*}[t!]
\caption[Propagation speed in a linear Poisson  network]{Mean Propagation speed in a linear Poisson  network with Aloha MAP $p$.
\label{table.ln}}
\begin{center}
\vspace{-2ex}
\begin{tabular}{|p{0.2\linewidth}|p{0.2\linewidth}|p{0.2\linewidth}|p{0.2\linewidth}|}
\cline{2-4} 
\multicolumn{1}{c|}{}  &isolated network & external noise or super-Poisson interference field & adding fixed
relays \\
\cline{2-4} 
\noalign{\vskip-10pt}
\multicolumn{4}{c}{}\\
\hline
\multicolumn{1}{|p{0.2\linewidth}||}{given distance $M$}  & rapidly increasing with $M$ and  stabilizing or decreasing back to $0$ for large $M$
  depending whether   $p<p_{\text{critical}}$  or not
                                & constant for  $M$ in a given interval,  approaching zero for  $M$ outside
  & increasing with $M$ and  stabilizing for large $M$\\  
  \hline
\multicolumn{1}{|p{0.2\linewidth}||}{asymptotically $M\to\infty$} & positive  provided \newline$p<p_{\text{critical}}$         & zero & positive for
  all $p$\\
   \hline
\end{tabular}
\end{center}
\vspace{-1ex}
{\footnotesize
super-Poisson field --- a field clustering more than Poisson in the sense of having larger Laplace transforms of
negative functions\\
$p_{\text{critical}}=\sup\{p: p\calD_1(p)<1, 0\le p\le 1\}$}
%\vspace{-2ex}
%\end{table}
%\begin{table}[h!]
\caption[Clustering paradox]{Clustering paradox; comparison of the impact of the external field of interfering nodes.\label{table.lni}}
\begin{center}
\vspace{-2ex}
\begin{tabular}{|p{0.3\linewidth}|p{0.3\linewidth}|p{0.3\linewidth}|}\cline{2-3} 
\multicolumn{1}{c|}{}&\multicolumn{1}{c|}{\hspace{1em}Poisson field\hspace{1em}\ }& \multicolumn{1}{c|}{\hspace{0.5em}Poisson-line    field \hspace{0.5em}\ }\\
\multicolumn{1}{c|}{}&\multicolumn{1}{c|}{(more homogeneous)}& \multicolumn{1}{c|}{(more clustering)}\\
\cline{2-3}
\noalign{\vskip-10pt}
\multicolumn{3}{c}{}\\
\hline
\multicolumn{1}{|p{0.3\linewidth}||}{capture probability}&\multicolumn{2}{c|}{$<$}\\\hline
\multicolumn{1}{|p{0.3\linewidth}||}{propagation speed} &\multicolumn{2}{c|}{$>$}\\
\hline
\end{tabular}
\end{center}
\vspace{-2ex}
\end{table*} 

We have studied performances of end-to-end routing in MANETs,
using a linear nearest-neighbor routing model  embedded in
an independent planar field of interfering nodes using Aloha MAC.
We have developed  numerically
tractable expressions for  several performance metrics such as the end-to-end
delay and speed of packet progression. 
They show how network performance can be optimized by tuning  Aloha and
routing parameters. In particular,
we show a need for a well-tuned lattice structure of wireless relaying
nodes (not back-boned) which help to relay packets on long random routes in the
presence of a non-negligible noise. 
We have also proved and explained why 
for a given intensity of interferers, the propagation speed is higher
when the interferers are more homogeneously distributed, even if, paradoxically, the probability of a single
successful transmission is smaller in this case. Numerical comparisons  in this regard have been shown for Poisson
and Poisson-line processes. The qualitative summary of the results is presented in Tables~\ref{table.ln}
and~\ref{table.lni}.

The strength of the present paper is that it offers exact, closed-form expressions 
of the performance metrics considered.
This raises the question of how far this rigorous approach can be extended. 

Although our analysis does not  take into account
packet queuing at the relay nodes, it
covers the practical case 
of broadcasting of emergency packets,  which are
transmitted in priority at each relay.
Our developments use Aloha rather than a CSMA 
protocol, which is generally more efficient than Aloha. 
Thus, our results can be considered as  conservative 
bounds regarding real VANETs
with a CSMA protocol.

To handle CSMA protocols, 
approximations will be certainly necessary since, in contrast to 
Aloha, the transmission by one node depends on the activity of
other nodes. Packet queuing further strengthens  this dependence. 
The paper does not consider mobile nodes, nor 
nodes leaving the network, which could be a more realistic
assumption regarding VANETs.

\addtocounter{section}{1}
\addcontentsline{toc}{section}{References} 
%\pdfbookmark[0]{References}{References} 
{\small  \bibliographystyle{unsrtnat}%{plainnat} 
\bibliography{linearNN}}

\end{document}